\documentclass[a4paper]{article}

\usepackage[english]{babel}
\usepackage[utf8x]{inputenc}
\usepackage[T1]{fontenc}

\usepackage[a4paper,top=3cm,bottom=2cm,left=3cm,right=3cm,marginparwidth=1.75cm]{geometry}

\usepackage{amsmath}
\usepackage{graphicx}
\usepackage[colorinlistoftodos]{todonotes}
\usepackage[colorlinks=true, allcolors=blue]{hyperref}

\usepackage{siunitx}
\usepackage{caption}
\usepackage{subcaption}

\usepackage{mathabx}

\title{Economic analysis of tidal stream turbine arrays: a review}
\usepackage{authblk}
\author[1]{Z. L. Goss}
\author[2]{D. S. Coles}
\author[1]{M. D. Piggott}
\affil[1]{Department of Earth Science and Engineering, Imperial College London, SW7 2AZ, UK}
\affil[2]{School of Engineering, Computing and Mathematics, University of Plymouth, Plymouth PL4 8AA, UK}
\date{\vspace{-5ex}}

\begin{document}
\maketitle

\begin{abstract}

This tidal stream energy industry has to date been comprised of small demonstrator projects made up of one to a four turbines. However, there are currently plans to expand to commercially sized projects with tens of turbines or more. As the industry moves to large-scale arrays for the first time, there has been a push to develop tools to optimise the array design and help bring down the costs. This review investigates different methods of modelling the economic performance of tidal-stream arrays, for use within these optimisation tools. The different cost reduction pathways are discussed from costs falling as the global installed capacity increases, due to greater experience, improved power curves through larger-diameter higher-rated turbines, to economic efficiencies that can be found by moving to large-scale arrays.  
A literature review is conducted to establish the most appropriate input values for use in economic models. This includes finding a best case, worst case and typical values for costs and other related parameters. The information collated in this review can provide a useful steering for the many optimisation tools that have been developed, especially when cost information is commercially sensitive and a realistic parameter range is difficult to obtain.  

\end{abstract}

\tableofcontents

\section{Introduction \& Motivation}\label{sec:intro}

Demand for sustainable and reliable energy sources is increasing, and tidal energy could prove an important part of the future energy generation mix, due to its predictable nature. However, tidal energy is a nascent industry compared to more established renewables such as wind and solar. These industries have seen significant costs reductions due to subsidies helping them learn from expensive early demonstrator arrays. These initial subsidies allowed wind and solar to progress to large-scale energy farms with a much lower price of energy. At the time of writing, the UK's largest onshore wind farm now consists of 215 turbines (Whitelee Windfarm) and the largest offshore farm has 189 (the Walney Extension). The maturity of the industry and the economies gained from large numbers of devices has led to the UK seeing a record low price per MWh from offshore wind of £39.65 in their third Contracts for Difference auction, in September 2019, for less established renewable technologies \cite{2019ContractsResults}.

In order to compete with more established technologies, tools must be developed to predict and help reduce the cost of tidal energy. In order for academically produced tools to be useful for industry, they need to be combined with the economic models and metrics that tidal investors are interested in. The inputs to these models are typically commercially sensitive and are therefore rarely publicly available. While exact values may only be used internally within industry, estimates of the inputs are useful for testing the academically developed tools. 

Previous design optimisation studies, such as \cite{Funke2016DesignApproach, Funke2014TidalApproach, Culley2016IntegrationArrays,duFeu2017TheProblem}, have optimised not only an array's design, but also the number of turbines in an array. This can be important in order to take advantage of the cost reductions that can be achieved with larger scale arrays, while also balancing against the diminishing returns as the average power generation per device falls with additional turbines. This review formats the cost inputs so they can be used in an economic model with a varying number of turbines and therefore can be used in the functional for a model that optimises both the location and the number of turbines. 

This review attempts to summarise the economic models most commonly used by the tidal energy industry and associated academia and to collate the publicly available cost information to produce appropriately formatted estimates for the inputs into these models. Section \ref{sec:econ_metrics} outlines the most commonly used economic metrics and their advantages and disadvantages. Section \ref{sec:cost_reduction} describes the different mechanisms expected to lead to the reduction in the price per MWh of tidal stream energy. Section \ref{sec:input_est} combines publicly available cost information, yielding data in a format useful for the input into the models described in Section \ref{sec:econ_metrics}. Finally, Section \ref{sec:conclusion} discusses the uses and the limitations of the information collated in this review. 

\section{Review of economic metrics for evaluating array design}\label{sec:econ_metrics}

When designing a tidal turbine array the choice of functional to optimise  can have a significant impact on the resultant array design. The following sections outline common metrics that are used for evaluating the performance of a tidal array, which could be used as the functional. 

\subsection{Power}\label{sec:power_alone}

Many papers \cite{Funke2014TidalApproach, Goss2018CompetitionRace,Divett2013OptimizationMesh} optimise array design for power alone, such that the functional is $J = P_{\text{avg}}$.
This can be an effective method of determining a suitable layout, especially if the size of the array is already specified, such as in \cite{Funke2014TidalApproach, Divett2013OptimizationMesh} In such cases the costs are relatively fixed, except for the costs that depend on the individual turbine locations, such as the cabling \cite{Culley2016IntegrationArrays} or distance and depth related installation costs, which are discussed in greater detail in Section \ref{sec:location_costs}. However, in general maximising for power is a reasonable proxy for maximising the economic performance. 

Problems in using the power alone as a functional arise, in particular, when the number of turbines is also allowed to vary and becomes a free parameter within the array design optimisation. Earlier work \cite{Goss2018CompetitionRace,Goss2019EconomicArrays} has shown that, when optimising for power alone, the optimal design will feature an impractically high number of turbines, where the overall power generation is at its maximum, but the capacity factor of the array, and therefore its profitability, is relatively low.
The optimisation algorithms will keep adding turbines, which slow the flow velocity through the site, until a point is reached where the blockage is so high that adding any more turbines will decrease the overall power generation. However, the extracted power per additional turbine will diminish long before the threshold where the overall power decreases. Given the high cost of installing devices the optimal economic performance will be achieved at a far lower number of turbines than the optimal power.

\subsection{Break even power}\label{sec:BEP}

As established in earlier work \cite{Goss2018AnRace,Goss2018CompetitionRace, Goss2019VariationsCosts}, adding additional turbines to an array may increase the overall power, but there comes a point where there are diminishing returns for each extra turbine installed, for example due to global blockage effects as well as the array being forced to expand into lower flow areas \cite{Culley2016TheArrays}.
From a financial perspective it may be more advantageous to have a smaller array where the average power per device ($\text{PPD} = \text{Avg Power} / n_t$) is higher, because tidal turbines are relatively costly to manufacture and install. In recent work \cite{Goss2019EconomicArrays,Goss2019VariationsCosts} a break even power, $P_{\text{BE}}$, is included in the optimisation functional to account for this problem and ensure that turbines are only added if they can generate enough power to cover their installation costs. Varying this parameter changes the trade-off priorities between the objectives of maximising power generation and minimising costs. 

The break even power is the average power over all turbines that needs to be generated in order for the array to break even over its lifetime. A formula for choosing a suitable break even power can be found by first using a simplified model for the expected cash flow as the functional, such that

\begin{equation} \label{eq:cashflow}
\begin{aligned}
J &= \text{Revenue} - \text{Cost} \\
&= \sum_{i=0}^{L} ( P_{\text{avg}} \times t_i \times T_e ) - \sum_{i=0}^{L} \text{Ex}_i ,
\end{aligned} 
\end{equation}
where $i$ is the year the costs are being evaluated over, 
$L$ is the lifetime of the array in years, $P_{\text{avg}}$ is the average power generated by the whole array in \mbox{MW}, $t_i$ is the number of hours of generation in year $i$, $T_e$ is the electricity tariff, i.e. the price per MWh the electricity generated is sold at, and $\text{Ex}_i$ is the sum of all array expenditure incurred in year $i$. 
If we assume $P_{\text{avg}}$ is independent of the year we can use this to find a critical average power per device that must be generated by the whole array to break even:

\begin{equation} 
P_{\text{BE}} \times n_t \sum_{i=0}^{L}  ( t_i \times T_e  - \text{Ex}_i ) = 0 ,
\end{equation}
\begin{equation}  
\iff P_{\text{BE}} = \frac{\sum_{i=0}^{L}  \text{Ex}_i}{ n_t \times \sum_{i=0}^{L} ( t_i \times  T_e )} . 
\end{equation}

If we assume that total expenditure is just a multiple of the number of installed turbines, $\text{Ex}_{i} = n_t \times \text{Ex}_{t,i}$, then a constant value for $P_{\text{BE}}$ can be found which is independent of array design:
\begin{equation}  \label{eq:bepdef}
P_{\text{BE}}  = \frac{\sum_{i=0}^{L}  \text{Ex}_{t,i}}{ \sum_{i=0}^{L} (t_i \times T_e)}. 
\end{equation}

In practice, $P_{\text{BE}}$ would probably decrease for larger scale arrays, due to economies of volume \cite{Smart2018TidalBenefit}. More sophisticated models to account for this are discussed below. Since the functional defined in \eqref{eq:cashflow} is invariant with respect to scaling, it can be seen that optimising with respect to expected cash flow is equivalent to optimising a functional of the form 

\begin{equation} 
J =  P_{\text{avg}} - \frac{\sum_{i=0}^{L}  \text{Ex}_{t,i}}{ \sum_{i=0}^{L} (t_i \times T_e)},
\end{equation}
which combined with the aforementioned simplifying assumptions can be reduced to 

\begin{equation} \label{eq:bep_func}
J =  P_{\text{avg}} - P_{\text{BE}} \times n_t. 
\end{equation}

Therefore for the array to generate a profit the average total power generated by the array must be more than the break even power times the total number of devices. If an appropriate break even power is chosen to reflect all of the costs in \eqref{eq:cashflow}, this choice of functional effectively maximises the profit. It penalises the addition of turbines which do not generate sufficient power to offset their costs, or which due to hydrodynamic changes lead to reductions in the yield of other turbines. 

An advantage of using the functional described in \eqref{eq:bep_func} is that it incorporates financial considerations to the optimisation through the use of just one simple variable, $P_{\text{BE}}$. However, this simplification means that the optimisation process has no flexibility to changes in strike price, annual yield variability, maintenance costs and hours downtime throughout the many years an array is operating for. It also excludes the effect of economies of volume, where larger-scale arrays would require a lower $P_{\text{BE}}$ to break even\cite{Smart2018TidalBenefit}. Both these limitations mean that, in reality, break even power would be a function of time and the size of the array, rather than a constant as assumed here. This might be an acceptable simplification if the optimal design was robust to changes in $P_{\text{BE}}$ that arise from these factors; however, previous studies \cite{Goss2019VariationsCosts} have shown the optimal design to be very sensitive to even small changes in $P_{\text{BE}}$.

Other papers have used similar or equivalent metrics to the break even power, with varying terminology. Iyer et al. \cite{Iyer2013VariabilityKingdom} assessed the viability of tidal sites around the UK and used the capacity factor as a simplified indicator of the economic performance of a site, on a per MW basis. They found an average capacity factor of 29.9\% (but ranging between 23.3 and 43.6\%) across all the sites they investigated in the UK. They noted that this metric has limited usefulness in models where turbine specifications chosen (such as diameter and rated power) are generic rather than tailored to the site being evaluated, therefore under-utilising the resource. 

Funke et al. \cite{Funke2016DesignApproach} first coined the term break even power, but also called it a cost coefficient. Their definition was very similar to the one given in \eqref{eq:bepdef}, except instead of evaluating about the break even point, they assumed a profit margin, $m$, was required such that
\begin{equation} \label{eq:profit_margin}
m = \frac{\text{Revenue} - \text{Cost}}{\text{Revenue}}
\end{equation}
and they made the assumption $m=73\%$. They assumed that the 73\% profit margin could be achieved by a 20m diameter turbine if the peak speed was \SI{3.5}{\m\per\s}, and used this to estimate a break even power of \mbox{452 kW}. They only tested the optimisation for this one break even power value, but noted that the break even power chosen was expected to have a large impact on the optimal number of turbines. 

Funke et al. \cite{Funke2016DesignApproach} also noted that the break even power model does not take a discount rate into account. Therefore no financial distinction is made between capital expenditure, i.e. upfront investments such as turbine and installation costs, versus operational expenditure, i.e. maintenance costs incurred years into the future. 

A further limitation of using break even power as the functional is that it makes too many simplifying assumptions and has no way to account for economies of scale. Below we discuss an adaptation to the functional to account for this. Overall, break even power represents a good early measure while insufficient financial information is available since all you need to know is an aspirational capacity factor. For example, if the developer's financial model finds that an array of \mbox{2MW} turbines needs to achieve a 40\% capacity factor to be profitable, then the break even power needs to be set to $2000\times0.4 =800$\mbox{kW}. However, once a more complete model of the costs is available a more detailed metric should be used that accounts for all the effects ignored by the above assumptions.

\subsubsection{Break even power with economies of volume}\label{sec:BEP_EV}

When the costs are assumed to increase linearly with the number of turbines, the break even power is independent of the number of turbines and therefore can be used as a constant value, as given in \eqref{eq:bepdef}. 
In practice that relationship will not reflect the true expenditure and the break even power will likely decrease as $n_t$ increases due to economies of volume. There are a number of ways this impact can be modelled, the simplest of which is to linearly decrease the break even power with the number of turbines, such that $P_{\text{BE}:\text{EV}}= P_{\text{BE}}-EV\times n_t$, where $EV$ is a coefficient for economies of volume, and $EV \ll P_{\text{BE}}$. This change can be made to the functional \eqref{eq:bep_func} so that

\begin{equation}  \label{eq:esbepdef}
J= P_{\text{avg}} - (P_{\text{BE}}-EV\times n_t)\times n_t = P_{\text{avg}}-P_{\text{BE}}\times n_t + EV\times n_t^2. 
\end{equation}

Using a functional of this format is equivalent to the expenditure instead decreasing quadratically with the number of turbines, such that:

\begin{equation}  \label{eq:constbepdef}
P_{\text{BE:EV}} 
= \frac{\sum_{i=0}^{L}  \text{Ex}_{t,i} \times n_t - EV \times n_t^2}{ n_t \times \sum_{i=0}^{L} T_e \times t_i}
= \underbrace{\frac{\sum_{i=0}^{L}  \text{Ex}_{t,i} }{ \sum_{i=0}^{L} T_e \times t_i}}_{P_{\text{BE}}} 
 - \underbrace{\frac{\sum_{i=0}^{L}  \text{EV} }{ \sum_{i=0}^{L} T_e \times t_i}}_{EV} \times n_t . 
\end{equation}

\begin{figure}[h!]
\centering
\includegraphics[width=100mm]{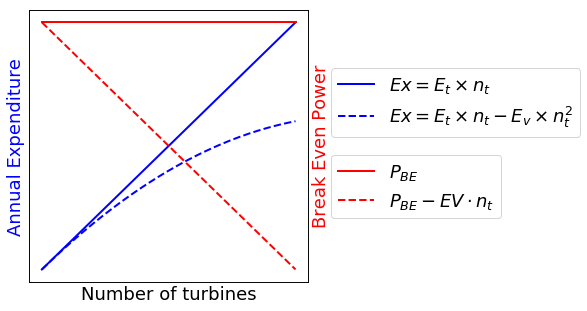}
\caption{A demonstration of how the relationship between expenditure per year and number of turbines in an array corresponds to a fixed or linearly decreasing break even power.}
\label{fig:expenditure}
\end{figure}

Figure \ref{fig:expenditure} shows how if the annual expenditure increases linearly with the number of turbines the break even power is fixed, however, if the expenditure declines quadratically with the number of turbines the break even power declines linearly with the number of turbines. Both rely on assumed relationships between the costs and number of turbines but are useful for initial investigations into how bringing costs and economies of volume into the functional effect the optimal array design. 

The impact of varying both the $P_{\text{BE}}$ and $EV$ parameters are explored in  \cite{Goss2019VariationsCosts}. It was found that as $P_{\text{BE}}$ falls, for example due to costs falling in the tidal industry with experience gained, the optimal array design features more turbines and generates more power overall, but the average power per device drops. It was shown that optimal array properties were more sensitive to changes in the region $300kW \leq P_{\text{BE}} \leq 500 kW$. This was even more notable when including economies of volume too; changing the value of $EV$ had little to no impact on the optimal design for very high or very low values of $P_{\text{BE}}$. For $300kW \leq P_{\text{BE}} \leq 500 kW$ however, it was seen that as $EV$ increased the total power generated and number of turbines increased, and the power generated per device decreased. 

\subsection{Net Present Value}\label{sec:NPV}

Equation \eqref{eq:cashflow} describes the sum of all cash flow across an arrays lifetime, however in practice this value is rarely used by investors. Long term investments, such as tidal arrays, rely on discounted cash flow (DCF) analysis to quantify the idea that money today is worth more than money tomorrow to investors. Current funds have the ability to earn interest and become worth more in the future, so investors need to find a way to adjust future cash flow to enable the comparison between costs and revenue in different time periods, in terms of their present day value. The Net Present Value (NPV) is a measure used to find the profit of a project is found by summing all incoming and outgoing cash flows per year, i.e. the revenues minus the costs from each year $i \in [0,L]$, adjusted according to the time value of money by a discount rate, $r$, such that 

\begin{equation} \label{eq:NPV}
	\begin{aligned}
    \text{NPV} &= \sum^L_{i=0} \frac{\text{Revenue}_i - \text{Cost}_i}{(1+r)^i} \\
    &= \sum^L_{i=0} \frac{P_{\text{avg}} \times t_i \times T_e  - \text{Ex}_i}{(1+r)^i}.
	\end{aligned} 
\end{equation}

Using the NPV as the functional for tidal array optimisation has many advantages. Unlike the previous metrics it takes into account the depreciating time value of money, which is important to investors. It allows for far greater flexibility in the financial modelling than the previous metrics. For example, instead of basing analysis on a time averaged power generation, $P_{\text{avg}}$, it is trivial replace this with a time-varying generation. This can be used to model the impact that the power variation due to the 18.6 lunar nodal cycle \cite{Thiebot2020InfluenceFrance, Haigh2011GlobalLevels} has on the economic performance of the array. It could also be used to include the effects of the anticipated degradation of the turbine performance, for example due to biofouling and algae build up \cite{SongThePerformance}, and increase in required downtime, due to a higher number of faults with age, as commonly observed in the wind industry \cite{Faulstich2011WindDeployment, Ziegler2018LifetimeUK}. However using an average power generation is very useful while there is still a lot of uncertainty about the planned installation year and fault rates decades into the future. 

A disadvantage of using NPV as the functional for tidal array optimisation, is that it is hard to compare between projects of different sizes. 
A smaller tidal array may have a far higher profit margin but a much lower NPV than a large tidal array. It is hard to interpret directly from the NPV how successful a project is. 

\subsubsection{Break even analysis}\label{sec:break_even_analysis}

There are a number of different economic metrics which can be derived when applying break even analysis to the Net Present Value formula. Equation  \eqref{eq:NPV} can be set to zero in order to obtain a value for the so-called break even point of an energy project. Then the resulting equation can be solved for each of the input variables, to determine the parameter value that would need to be achieved for the project to break exactly even over its lifetime. Each of these parameters at break even can be seen as the minimum requirement, and any improvement on these will result in a positive NPV and the project will generate a profit.

The most common metrics are derived by solving for the strike price, to obtain the levelised cost of energy (LCOE), solving for the lifetime, to obtain the payback period (PP), and solving for the discount rate, to obtain the internal rate of return (IRR).  

\subsubsection{Discrete versus continuous discounted cash flow analysis}\label{sec:discrete_continuous_dcf}

All present value based calculations require summing all the cash flow over a lifetime and discounting them according to the time at which they occurred. The net present value calculation given in \eqref{eq:NPV} is setup so that the discount rate is applied on a yearly basis. It is possible to instead apply the discounting on different time intervals, e.g. semiannually, quarterly or monthly, or to use a natural log based formula to apply discounted cash flow analysis continuously. 

Using the discrete compounding method, the present value, $PV$, of a future cash flow, $FV$, is
\begin{equation} \label{eq:discrete_PV}
	\begin{aligned}
    \text{PV} = \frac{\text{FV}}{(1+\frac{r}{p})^{pi}}  ,
	\end{aligned}
\end{equation}
where $p$ is the number of compounding period per year (for example, $p=1$ for annual compounding or $p=4$ for quarterly compounding), $i$ is the number of periods into the future that the future cash flow occurs, and $r$ is the annual discount rate. 

By comparison the continuous compounding method has no period over which the compounding is applied and can instead be found from
\begin{equation} \label{eq:discrete_PV}
	\begin{aligned}
    \text{PV} = \text{PV} \times e^{-ri}  ,
	\end{aligned}
\end{equation}
where in this case $i$ is a continuous value for the number of years into the future that the cash flow occurs. 

A review into the use of discrete and continuous discount rates by Lewis et al. \cite{Lewis2009RealRates} found, in a literature review of finance and engineering economics journals, that the split between discrete and continuous discounting of future cash flows was fairly even. More mathematical papers tended to use continuous discounting, whereas more applied papers used discrete discounting. In practice many tidal array developers are not likely to have precise knowledge of when exactly costs will be incurred, especially not at the stage of designing an array. Therefore discrete modelling on an annual, or quarterly, basis is sufficient for this application. 

\subsection{Levelised Cost of Energy}\label{sec:LCOE}

The levelised cost of energy (LCOE) is a proxy for the average price of energy, $T_e$ [$\pounds$/MWh], that an array must receive to break even over its lifetime. It takes into account discounting and is calculated by finding the Net Present Value of the unit-cost of electricity over the lifetime of the array, by setting expression \eqref{eq:NPV} to zero and rearranging to find the price of energy;  

\begin{equation} \label{eq:LCOE}
	\begin{aligned}
    \text{LCOE} = \frac{\text{discounted cost}}{\text{discounted  energy}} = \frac{\sum_{i=0}^{L} \frac{\text{Ex}_i}{(1+r)^i}}{\sum_{i=0}^{L} \frac{P_{\text{avg}} \times t_i}{(1+r)^i}}.
	\end{aligned}
\end{equation}

LCOE will be focused on as the main metric for economic optimisation here since many studies and organisation use it as the most established form of estimating the lifetime costs of an energy generation project \cite{Vazquez2015DeviceEnergy,Vazquez2015LCOEEnergy, Ouyang2014LevelizedChina}. It is an effective benchmarking technique for the comparison of multiple energy generation technologies, and in this case multiple array designs. LCOE, IRR and PP are all effectively simplifications of NPV, where an input variable can be removed by instead only investigating the parameters required to break even. Often it is sensible to isolate the variable which has the most uncertainty in it. The price of energy $T_e$ can vary greatly depending on subsidies available to early stage renewable energies, so LCOE predictions often encapsulate less uncertainty than IRR and PP, because an assumption of $T_e$ is not required. 

The LCOE is the most commonly used approach for estimating the cost of energy over the lifetime of a project, for both tidal energy and other renewables \cite{Vazquez2015DeviceEnergy,Vazquez2015LCOEEnergy, Ouyang2014LevelizedChina, Smart2018TidalBenefit}. Since it is not as sensitive to array size as NPV or other metrics, it enables simple comparison across a range of different projects. 

\subsection{Payback Period}\label{sec:PP}

Another metric that can be derived from break even analysis of the Net Present Value formula is the Payback Period (PP), which is the lifetime array must operate for in order to break even. If the planned lifetime of the array is longer than this the array can be considered to be generating profit, if the lifetime of the array is shorted than this it will be making a loss.

The Payback Period can be found by calculating the NPV of every year from 0, until the NPV becomes positive. The critical year, $i_cr$, is the last year before the NPV becomes positive, i.e. the year before the project breaks even. A payback period can be calculated by estimating how far through the year the project breaks even;

\begin{equation} \label{eq:PP}
	\begin{aligned}
    \text{Payback Period} = i_{cr} + \frac{\text{NPV}_{cr}}{\text{NPV}_{cr} - \text{NPV}_{cr+1}}.
	\end{aligned}
\end{equation}

\subsection{Internal rate of return}\label{sec:IRR}

Finally, \eqref{eq:NPV} can be rearranged to find the discount rate at which the project breaks even, call the Internal Rate of Return (IRR). If the IRR is higher than the projected discount rate, then investors can anticipate a profit over the course of the array's lifetime.
IRR is harder to calculate because, when trying to find the discount rate at which a project breaks even, $r = \text{IRR}$, the formula can be expanded,

\begin{equation} \label{eq:IRR1}
	\begin{aligned}
    0 & = (E_0 \times T_e - \text{Ex}_0)(1+\text{IRR})^L + (E_1 \times T_e - \text{Ex}_1)(1+\text{IRR})^{L-1} + (E_L \times T_e - \text{Ex}_L),
	\end{aligned}
\end{equation}
but never to a form which can be solved analytically. 
However, a nonlinear solver such as the Secant method can be applied to find the IRR numerically. Each new iteration, at iteration number $n+1$, is found from the previous two estimates, such that

\begin{equation} \label{eq:IRR2}
	\begin{aligned}
   \text{IRR}_{n+1} = \text{IRR}_n - \text{NPV}_n \times \left( \frac{\text{IRR}_n - \text{IRR}_{n-1}}{\text{NPV}_n - \text{NPV}_{n-1}} \right).  
	\end{aligned}
\end{equation}
 
Solving this is more computationally expensive than finding the NPV, LCOE or PP. While the difference may be negligible if only performing the computation a small number of times for finalised array design, it may become a significant hindrance when used as the functional for a large number of optimisation and adjoint evaluations. 

\subsection{CAPEX versus OPEX}

The expenditures in each year, $\text{Ex}_i$, can be split up according to whether they occur before or after the array goes into production. Businesses typically describe all costs involved acquiring assets and setting up the business as Capital Expenditure (or CAPEX). Vasquez et al. \cite{Vazquez2016CapitalApproach} estimated that the capital costs for a tidal stream array break down as 41\% device costs, 26\% foundations costs, 15\% installation costs, 13\% cable costs and 5\% grid connection costs. By comparison, the MeyGen1A project found that the main contributors to the CAPEX were turbines (39\%), onshore balance of plant (BoP, the supporting components and auxiliary systems, 19\%), offshore works (13\%) and substructures (11\%) \cite{2020LessonsReport}. 

In the cash flow model described in \eqref{eq:cashflow}, the CAPEX would  typically be 
\begin{equation} \label{eq:CAPEX_def}
	\begin{aligned}
    \text{CAPEX} =  \sum_i \text{Ex}_i, \;  \forall i \;\;  \text{s.t.} \;\;  t_i = 0 , 
	\end{aligned}
\end{equation}
where $t_i$ is the number of hours of generation in year $i$ i.e. the costs incurred before the array is operational. In more complicated arrays the CAPEX may include the expenses associated with upgrading physical assets, for example in arrays which expand from small demonstrator arrays to larger scale farms, the installation costs are still considered to be CAPEX, even if they are incurred after the first turbines start generating. 
For simplicity, in this work it will be assumed that all the capital expenditures occur in year $i=0$ and that the array will start production in year $i=1$. This is an easy assumption to remove when applied to real tidal arrays, by developers who have full cash flow models of their anticipated costs. CAPEX will be represented in the following models as $\text{CA} \equiv \text{Ex}_0$.

The expenses of normal business operations are called the operational expenditure (or OPEX). OPEX is measured on an annual basis, here starting from year $i=1$ up to $i=L$, the operational lifetime of the array, and will be represented in the following models as $\text{O}_i$. Typically OPEX includes costs such as rent, payroll, insurance and maintenance. For a tidal farm this may include standard inspections, maintenance, repairs and costs of vessels and staff to perform these tasks. The MeyGen 1A project reported a £1.4m OPEX per year in its four turbine array, with its main components being lease and insurance (32\%), unplanned maintenance (21\%), planned maintenance (15\%) and spare parts (14\%) \cite{2020LessonsReport}.

The final costs to be considered in a tidal energy deployment are the decommissioning costs. These are the costs or removing the turbines, anchoring and cabling from the water and safely ending the operation. There are a number of different ways in which these costs may be covered - from upfront cash security to accrual or insurance\cite{2018MarineDevelopments}. Since these costs could either be incurred through an agreement as part of the CAPEX, or will be delayed to after the final year of operation $i=L+1$ it is not clear how to include these costs in an academic model at this time. If incurred in year $i=L+1$ the costs are likely to have very little impact on metrics based on net present value analysis because the cost will be heavily deprecated due to discounting. Due to minimal present value impact and the large amount of uncertainty about how this will be finances for large-scale tidal stream arrays and limited information to base estimates on how much it will cost, this is not included in our economic assessments at this point in time. 

\begin{figure}[!t]
\centering
\subfloat[][]{\centering\includegraphics[height=50mm]{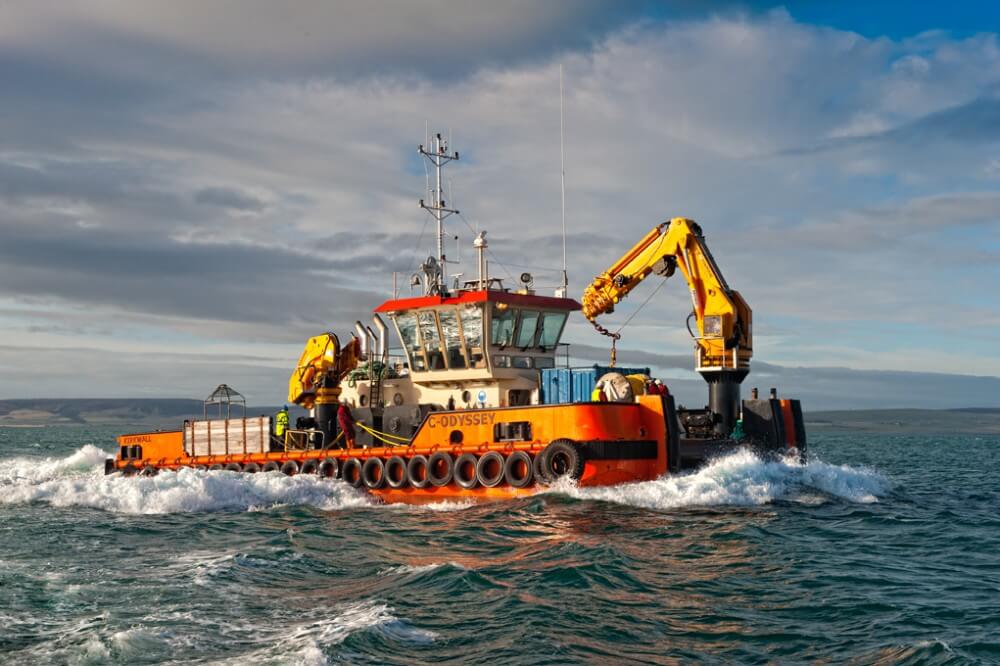}\label{fig:c-odyssey}} \
\subfloat[][]{\centering\includegraphics[height=50mm]{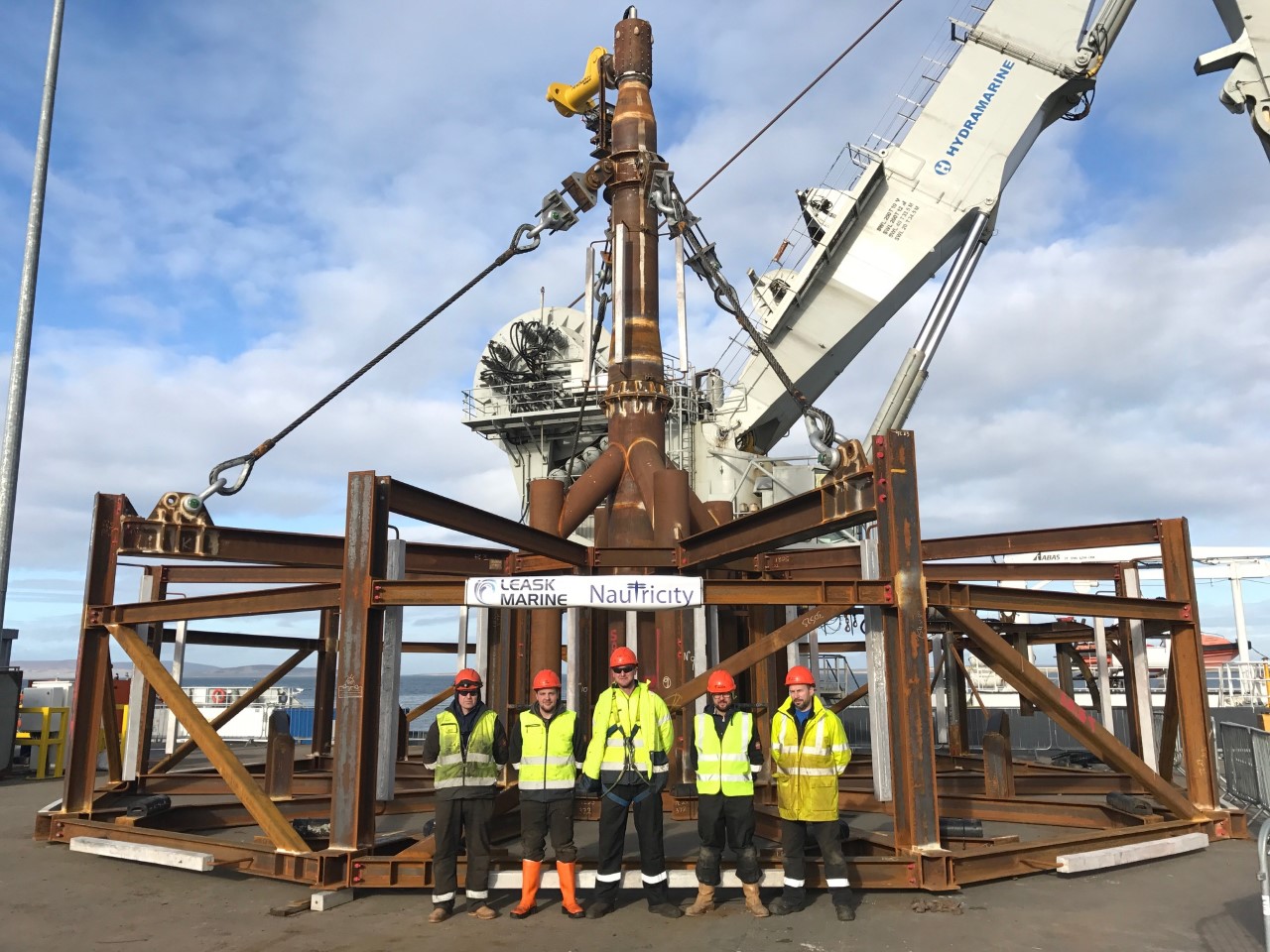}\label{fig:nautricity}} 
\caption{(a) The Leask Marine Ltd. MV C-Odyssey, an example of a multi-cat vessel used for installation and removal of turbines. (b) The Leask Marine Ltd. team after the successful removal of a Nautricity turbine's base with a multi-cat vessel.}
\label{fig:c-odessey}
\end{figure}

An investigation into decommissioning costs by Marine Scotland \cite{2018MarineDevelopments} found that many developers acknowledged that decommissioning had not been an explicit consideration on the design of devices, however a focus on reducing the cost of installation lead to designs which were inherently easier to remove. Technological advances which reduced the decommissioning cost as a by product included devices designed to be towed to (or from) location and devices that could be removed in modules for maintenance. Full decommissioning plans are not enforces by regulators at the marine licensing stage because the design may not be fixed. \cite{2018MarineDevelopments} found that the main cost driver is the cost of the vessels required to remove the infrastructure. Many tidal developers have designed devices to allow installation and removal with low cost multi-cat vessels, for example see Figure \ref{fig:c-odessey}. In demonstrator arrays material costs are low as many parts can be repurposed or stored. As arrays become larger the cost of dismantling and recycling the devices and foundations is likely to increase.

\cite{2018MarineDevelopments} estimated that floating tidal has a decomissioning cost of approximately £100,000 per device, with gravity-base foundations costing £200,000 per device and monopile foundations costing £500,000 per device. However, it is noted that these estimated could change substantially as design concepts and vessel rates change. Predicting these costs a couple of decades into the future leaves a lot of uncertainty for developers.

\subsection{Other metrics to consider}\label{sec:other_metrics}

\subsubsection{Environmental impact}\label{sec:env_impact}

The metrics described in the sections above all focus on the financial performance of a tidal array, however there are many other factors to consider. One of the most important, for ensuring that developments get consent to deploy, is minimising the negative environmental and ecological impact of the array. 

There have been numerous studies on environmental impact assessment in tidal and other ocean energies \cite{Ahmadian2012Far-fieldTurbines,Ahmadian2012AssessmentOutput,MccluskieBirdsReview}. The most significant effects to account for are damage to habitat and health of marine species, and sediment transport. While the problem of assessing the environmental impact of a tidal stream array is well researched, there are limited studies on how to optimise array design whilst accounting for both the economics and the environmental impact. A notable and novel example of using power and environmental impact for multi-objective optimisation is du Feu et al. \cite{duFeu2017TheProblem,duFeu2019TheApproach}. 

\cite{duFeu2017TheProblem} uses the extent to which an array alters the flow in the region as a proxy for the environmental concerns and demonstrates a method of finding a Pareto front where there is a trade-off between the conflicting objectives of maximising array yield while minimising the change in the hydrodynamics. This approach is useful for minimising the negative impacts on marine ecosystems because alteration of the tidal flow can affect the dispersion of propagules, material used by marine organisms to propagate between areas and a key part of many marine life-cycles,  \cite{Shields2011MarineEnvironment} and there are habitats which are sensitive to flow speed, direction and level of turbidity, for example scallop nurseries \cite{Eckman1989EffectsSay}. \cite{duFeu2017TheProblem} combined environmental impact assessment at the array design stage by maximising a functional of the form $J = w_p \times P(m) - w_i \times I(m)$, where $P(m)$ is the power generated by the set of turbine locations, $m$, and $I(m)$ is the impact of those turbines, measured as the array's effect on the ambient flow velocities. The relative importance of each term is modelled by $w_p$ and $w_i$, the power and impact weights, respectively. However, only a limited set of weighting were tested ($w_p=1$ and $w_i=0$, $w_p=0$ and $w_i=1$, $w_p=1$ and $w_i=1$) and realistic values for the weightings were not discussed. A further limitation of this approach is that the number of turbines was fixed, so varying the number of turbines would allow for greater exploration of the trade-off between power, and therefore profit, and environmental impact. 

\cite{duFeu2019TheApproach} extended upon the approach of \cite{duFeu2017TheProblem}, however it optimised the functional $J = w_p \times \text{Profit}(d) - w_h \times H(d)$, where $H(d)$ is a measure of habitat impact, $w_h$ is the relative importance of the habitat impact, and the turbines are instead modelled by a continuous density function, $d$, which allows for the number of turbines and their locations to be optimised simultaneously. $\text{Profit}(d)$ takes the same format as the break even power optimisation described in Section \ref{sec:BEP}.
The habitat impact, $H(d)$, was found through maximum entropy modelling, which was used to  generate habitat suitability maps for species that respond to changes in bed-shear stress. The maximum entropy was calculated using MaxEnt, an open-source habitat suitability model \cite{Phillips2013MaxEnt}, and was used to evaluate the impact of a tidal array on the distribution of a specific species at each timestep of the hydrodynamic model.
This study investigated the impact of different tidal arrays on two species, the  acorn barnacle (Balanus crenatus) and the brown crab (Cancer pagurus), because the former prefers higher flows speeds and tends to react negatively to reductions in bed shear stress and the latter prefers lower flow speeds and reacts positively to reductions in bed shear stress. 
The relative importance of habitat suitability compared to profit is modelled by the choice of weights of each corresponding term in the functional. The results were demonstrated on four example sites, including the Pentland Firth, location of the world's highest capacity tidal stream array, at the time of writing. While this paper demonstrated a useful method for incorporating the concerns of habitat suitability into the functional, each example site was only demonstrated for one species at a time, either the corn barnacle or the brown crab, and more work needs to be carried out to automate the combination of economics and environmental impact optimisation. 

\cite{Nelson2017AConstraints} compared the energy extraction of small-scale (five turbine) array designs that considered environmental constraints compared to those that do not. They found that taking into account environmental constraints decreased the overall power generation of the optimised array design slightly. However, they developed a flexible framework that allowed power maximisation while remaining within the bounds of easily defined constraints to changes in the flow and seabed stress. 

The present work does not include environmental impact in the economic analysis and optimisation methods discussed. As demonstrated by the studies discussed above, minimising the environmental impact often leads to a trade-off where the array yield, and therefore profitability, are reduced. It is acknowledged that further work should combine the economic optimisation methods with the environmental optimisation techniques outlined in works such as \cite{duFeu2017TheProblem,duFeu2019TheApproach,Nelson2017AConstraints}, but this would require overcoming the limitations of each study (for example by studying the impact of a broader range of species) and selecting weightings for the trade-off between optimal economics and environmental impacts. 

\subsubsection{Location based costs}\label{sec:location_costs}

In Section \ref{sec:BEP}, expenditure is assumed to be a function of the number of turbines, but in practice is it likely to depend on their location too. Two of the most significant factors in tidal energy costs that vary with location are the depth and the distance to shore \cite{Vazquez2016CapitalApproach}, and both have been demonstrated to be two of the main cost drivers in offshore wind \cite{Ederer2015EvaluatingApproach}. As both increase the installation and maintenance of the turbines become harder to carry out, and the costs increase. They also affect the environmental loads to which the turbines are exposed, the cable costs and electricity losses. 

There have been some studies which optimise the design tidal arrays based on the location dependent costs. On a macro-level Vazquez et al. \cite{Vazquez2016CapitalApproach} assesses the spatial distribution of capital costs, coupled with a Navier-Stokes flow solver, to balance the capital investment against the energy productions and find a map of the LCaOE (levelised capital cost of energy). This is used as a decision parameter to identify the best sites to install tidal stream arrays. This can be done as a precursor to the array optimisation methods where the number of turbines and their individual locations are optimised. 

Studies in which the micro siting of tidal stream arrays is optimised with respect to location-varying costs include \cite{Sullivan2013OptimisationAlgorithm} and \cite{Culley2016IntegrationArrays}. \cite{Sullivan2013OptimisationAlgorithm} and \cite{Culley2016IntegrationArrays} focus on optimisation of a given number of turbines, where the functional is based on an approximation to profit. \cite{Sullivan2013OptimisationAlgorithm} includes a model for cable costs per metre and support structure costs that depend on the nature of the seabed, the depth and the peak moment resistance required. \cite{Culley2016IntegrationArrays} optimises the income as a function of power minus the cost as a function of cable length. Both use genetic algorithms to minimise the cable length. 

While important, the costs that depend on the individual turbine locations make up a smaller variation in the overall costs than those due to changes in the number of turbines. Vasquez et al. \cite{Vazquez2016CapitalApproach} estimated that the cabling costs make up 13\% of the CAPEX for a typical tidal stream array, whereas it accounted for 9\% of the CAPEX and none of the OPEX breakdown reported by MeyGen Phase 1A \cite{2020LessonsReport}. Culley et al. demonstrated an optimisation of a tidal array in Orkney, where they optimised for power alone, maximisation of financial return including cable costs, as well as the maximisation of financial return where the cost of the cable per unit length is doubled; in each case the returned cable lengths of the optimal arrays were found to be 9.70km, 9.23km, and 8.71km respectively \cite{Culley2016IntegrationArrays}. So even when optimising for an exaggerated cable cost, including the cabling in the functional in this case only reduced the cabling cost by around 10\% and therefore the overall cost in the order of 1\%.

Optimisation while accounting for location varying costs can be very computationally expensive, and usually comes at the sacrifice of other modelling capabilities. For example, \cite{Sullivan2013OptimisationAlgorithm} uses a genetic algorithm to optimise the locations while taking into account variations in cable and support structure costs with location, but only within the framework of a relatively simple 2D wake deficit based model, and only for uniform flow in one direction.
Optimisation of the cable routes can produce savings that are significant, but are a lower priority compared to optimising the balance between power and number of turbines. It is therefore suggested in the following work that location varying costs need not be included as another variable in very computationally expensive micro-siting optimisation. Instead location based costs can be taken into account either at the macrositing stage, as demonstrated by \cite{Vazquez2016CapitalApproach}, through exclusion zones, where a maximum depth or steepness requirement is enforced \cite{Schwedes2017MeshOptimisation},
or post micro-siting, for example where a genetic algorithm or travelling salesman type problem  \cite{Culley2016IntegrationArrays} is used to minimise the cable length for a given set of turbine locations.

Furthermore, currently there is not enough publicly available information to build a model that can predict the costs as a function of different distances to shore, depths and number of turbines, so this paper just focuses on varying the latter. Furthermore the sites selected for first-generation tidal turbine arrays are usually a relatively short distance to shore and require depths of 25--50m \cite{Iyer2013VariabilityKingdom,Lewis2015ResourceArrays}. It is anticipated that later generations of tidal stream will be designed for operation in deeper water, where modelling of location-varying costs will become more crucial \cite{Iyer2013VariabilityKingdom}. 

\section{Cost reduction pathways}\label{sec:cost_reduction}

Estimates from the Offshore Renewable Energy Catapult (OREC) in 2018 anticipated that tidal stream energy costs in the UK could come down by 70\% from a representative LCOE of £300/MWh to £90/MWh as the installed operational capacity rises from \mbox{10MW} to \mbox{1GW} \cite{Smart2018TidalBenefit}. Part of this reduction in costs is due to saving time and cost through ``learning rates'', part is due to ``economies of scale'', such as moving to larger rotor and higher rated devices, and part is due to the move to larger scale arrays which benefit from ``economies of volume''. 
It is important to be aware of the distinction between the different types of cost reduction, because they present opportunities at different stages of array development to cut to cost of energy. Learning rates can only be exploited by delaying the deployment of an array to avoid the first mover disadvantage, and the latter two can be exploited by choosing the optimal turbine specification and the optimal number of turbines respectively. 
The following work using the same terminology of each type of cost reduction as the OREC report on tidal stream and wave energy cost reduction \cite{Smart2018TidalBenefit} and a study by Coles et al. on mechanisms for reducing the cost of tidal stream energy, applied to the context of the expansion of the MeyGen tidal array from currently operational Phase 1A to the plans for Phase 1C \cite{Coles2019MechanismsEnergy}. 

\subsection{Learning rates}

Learning rates encompass all cost reductions where the costs of an industry fall with time due to increased knowledge. This will not be included in the following models as we assume we are optimising arrays for developers which have a fixed level of experience. However, it is worth considering that our bounds for costs will likely become overestimates with time, and that sources for typical values to use in the following models will be more reflective of the current state of the industry if more recent. The extent to which learning rates will impact costs can be estimated by observing what has happened in other industries, such as wind power, where costs are decreasing as the cumulative installed capacity increases. They are typically represented by a percentage reduction in costs with each doubling in cumulative capacity. 
An Arup produced `Review of Renewable Electricity Generation Cost and Technical Assumptions' estimates that the doubling of installed tidal capacity will result in a 13\% CAPEX cost reduction and a 19\% OPEX fall \cite{Arup2016Department}, whereas OREc assumed 13\% for CAPEX but a more conservative learning rate of and 11\% for OPEX \cite{Smart2018TidalBenefit}. 
The impact of learning rates are most significant in the near term for fledgling industries like tidal, because when the cumulative installed capacity is so low it is easier for it to double. Learning rates can be accelerated by collaboration, but hindered by protection of intellectual property preventing these opportunities. Cost reduction through learning can be grouped into two main categories; learning-by-doing and learning by innovation.

Learning-by-doing cost reduction relates to project developers repeating processes with each array that they install, thus learning how to optimise procedures and minimise costs. Examples of cost reductions through learning include better planning of operations and maintenance using operational and weather data, supply chain optimisation and automation, and familiarity with installation sites \cite{Smart2018TidalBenefit}. Standardisation of foundation and component design will become possible as the number of devices deployed increases, so costs for the industry will fall. 

Repeat installations of tidal arrays can also lead to improved proof of concept, resulting in increased confidence from investors and a reduction in the cost of capital as projects are seen as less risky. The cost of capital is the required return for a project to be worthwhile, for internally financed projects it is the cost of equity (mostly comprising of dividends), and for externally finance projects it is the cost of debt (the effective interest rate paid on debts). The weighted average cost of capital (WACC) is used to combine the cost of equity and debt into one figure and it is often used as the discount rate for present value analysis of future cash flows, such as the NPV and LCOE models in Equations \ref{eq:NPV} and \ref{eq:LCOE} respectively. 
Currently tidal projects are financed through a combination of grant support and private finance, and tidal has a far higher cost of capital than the more established offshore wind industry. Increased confidence from financial institutions will see a rise in financing from commercial debt too. 
OREC estimated the cost of capital for 10MW cumulative capacity is 10\% and that it will fall to 8.4\% by 100MW, 8.0\% by 200MW and 7.1\% by 1GW \cite{Smart2018TidalBenefit}. They believe that these reductions can be achieved through increasing the proportion of debt finance (which has a lower interest rate of around 4.5\%) and reducing the rate of equity premiums (from around 10\% to 8\%). The more familiar and stable the technology is perceived to be, the greater the cost of capital reductions that can be obtained. OREC noted that each 1\% reduction in WACC (and therefore discount rate) resulted in a 6\% reduction in LCOE in their studies. 
MeyGen released their lessons learnt report from the 1A construction phase \cite{Ltd2017LessonsCONTENTS} which found that the turbines were exceeding their contractual key performance indicators such as average power coefficient (8\% increase), capacity factor (20\% increase) and lifetime energy yield estimate (18\% increase). These substantial improvements demonstrate a reduced risk to investors and could lead to a reduction in the cost of capital for future tidal arrays. 

Learning by innovation covers cost reductions due to technological improvements such as improved performance and reliability of individual components. Some of these innovations may arise from the experience gained in the more mature offshore oil, gas and wind industries, and will therefore reduce the level of perceived risk to investors. 
Innovation may reduce the operational costs or improve the structure or the availability of the turbines. For example, a consortium led by ITPEnergised, has designed a customised barge for installing turbines which may allow the industry to move away from hiring expensive dynamic positioning vessels. These barges are anticipated to be much cheaper for installation of larger scale arrays \cite{DynamicallyITPEnergised}. Leask Marine Ltd have successfully used multi-cat vessels to install and remove multiple tidal turbines and their subsea bases, demonstrating the potential for multi-cat vessels to replace the need for multi-million-pound large dynamic positioning drilling ships and jack-ups.  
Improved electrical connectors are also expected to reduce the cost of tidal in the short term. 
Wet mate connectors are a technological development that allow the cable connections to the turbines to be made sub-sea rather than on the vessel, thus simplifying the installation on operations. They allow turbines to be installed onto their foundations in less than 60 minutes, and reduce the installation cost by 65\% compared to a turbine with a dry mate connector where the export cable must be brought onto the vessel to connect it to the turbine \cite{Coles2019MechanismsEnergy}.
Again the following models will not include learning by innovation, because this work is aimed at developing a method to assist developers in optimising array design given the technology options currently available to them. 

\subsection{Economies of scale}

The following work uses the same distinction between the economies of scale and volume as \cite{Smart2018TidalBenefit, Coles2019MechanismsEnergy}. Economies of scale is used to exclusively refer to the cost reductions gained from moving to larger rotor, higher wattage turbines (i.e. increased scale of the turbines themselves) and not to the effects of increasing the number of turbines. 

Coles et al. \cite{Coles2019MechanismsEnergy} demonstrated the significant impact that economies of turbine scale could have, by investigating the predicted yield of the 1.5MW rated AR1500 device compared to the 2MW rated AR2000. The AR1500 is currently operational as part of the MeyGen Phase 1A, the AR2000 is the next generation device by SIMEC Atlantis Energy, which is expected to be deployed in future phases of the MeyGen project. Both devices are capable of accommodating a range of physical options, but Coles et al. assessed an 18m diameter AR1500 with a hub height of 14m to a 20m diameter AR2000 with a hub height of 15m. 
Coles et al. \cite{Coles2019MechanismsEnergy} estimated that economies of turbine scale could lead to a reduction in LCOE of 17\%, 20\% or 23\% due to the 29\% uplift in anticipated yield when progressing from the AR1500 to the AR2000, depending on whether the CAPEX increases by 10\%, 5\% or 0\% between the two devices. It is difficult to predict how much CAPEX will increase between the turbine designs, because the costs are commercially sensitive. They are likely to increase due to higher loading from larger blades, increased generator size and hub height, but the AR1500 was designed to meet conservatively high loads, and through the proof of concept in MeyGen 1A, the AR2000 may not require as conservative a design. However, for each of the CAPEX scenarios investigated, it was shown that small achievable changes to the turbine specification can result in significant near-term reductions to the LCOE of tidal energy. 

Foundation costs also make up a significant proportion of the lifetime costs of an array, so some developers plan to use several rotors on the same supports to spread the foundation costs over a higher rated power, and achieve economies of scale \cite{2013OceanM}. 

\subsection{Economies of volume}

Economies of volume are found when costs can be spread across more turbines in an array; when the installed capacity in \mbox{MW} for a potential tidal site is increased, the effective cost per \mbox{MW} decreases. For example, the cost of mobilising (preparing the vessel) and demobilising (unloading and returning of the vessel) for offshore operations, installations and maintenance falls per turbine as the number of turbines increases. For small arrays these activities take up a significant proportion of the total vessel time. \cite{Coles2019MechanismsEnergy} demonstrated that in MeyGen 1C the anticipated number of mobilisation and demobilisation days per turbine falls from 1.13 to 0.83 if the number of turbines increases from the current 4 to a planned 36. 

There are fixed costs in tidal array development, such as the site evaluation and substation costs, which lead to a reduced cost per MW when they are split across a greater number of turbines. Potential cost reductions due to economies of volume can also include bulk order discounts, reduced production costs per unit and savings due to serial production and standardisation of common components. For example, dedicated manufacturing facilities for large-scale turbine orders will be more cost-effective, but this is not possible for smaller arrays. Also, inter-array cable costs may increase approximately linearly with the number of turbines but the costs of a substation and the much more expensive export cables to shore will be roughly fixed, so the more turbines generating power, the lower the total cable cost per \mbox{MW}.

To date tidal stream has only been demonstrated in arrays with small numbers of turbines; the world’s first arrays being three turbines by Nova Innovation off the coast of Shetland and four turbines installed by Meygen in the Inner Sound of the Pentland Firth. 
Even without cost reductions due to experience or improved technology, substantial cost reductions could be seen in the immediate term, just by moving from these demonstrator sized arrays to commercially sized ones. 

It is worth noting that while costs are expected to fall with time and volume, there are more significant cost reductions to be made while tidal energy is a relatively new industry. The cost reductions as the tidal industry advances from small demonstrator arrays to the first large-scale commercial arrays will be significant but as the industry develops the potential to reduce costs will diminish. 
Some aspects of tidal arrays are already well-established, building on what can be learnt from other offshore energies. For example, electrical connection to the grid (which makes up around 5\% of the lifetime costs of tidal arrays) has less potential for dramatic cost reduction due to developments already being carried out by the offshore wind industry, so the learning reductions have to some extent plateaued \cite{2013OceanM}. However, there is still potential for significant economies of volume to be achieved on these costs through subsea hubs which will allow the electrical connection of multiple devices resulting in cheaper configurations. 

\subsubsection{Expenditure break down}

The capital costs are split into fixed costs and turbine-dependent costs, to help include economies of volume into the model. 
There are some costs that need to be overcome no matter how many turbines are installed and some that increase linearly with number of turbines. For example, the inter-array cabling may be considered a turbine-dependent cost because the length of cables will be proportional to the number of turbines. However, the export cable will be a fixed cost, since there will always need to be a cable that transports power generated from the substation to shore, regardless of number of turbines. 
The CAPEX can therefore be written as

\begin{equation} \label{eq:CAPEX}
	\begin{aligned}
    \text{Ex}_{i=0} = \text{CAPEX} = \text{CA}_{f}+\text{CA}_{t}\times n_t  . 
	\end{aligned}
\end{equation}

Similarly the operational expenditure are the costs incurred every year after installation and are assumed to linearly increase with number of turbines. 
Furthermore, this study assumes that the OPEX is the same year on year, for simplicity, however developers could use this model with costs that vary very easily. For example, some components of the array may need maintenance every year whereas others may need maintenance every few years, causing a routine increase in costs. 
By splitting the OPEX into fixed and turbine dependent costs it can be written as

\begin{equation} \label{eq:OPEX}
	\begin{aligned}
    \text{Ex}_{i>0} = \text{OPEX} = \text{O}_{f}+\text{O}_{t}\times n_t  . 
	\end{aligned}
\end{equation}

\subsection{Revenue}\label{sec:revenue}

In order to build a complete financial model, income must be accurately predicted, as well as expenditure. For a tidal farm, income depends primarily on two factors -- the net energy output and the tariff at which this energy is sold to the grid (which may well be higher than the market price due to subsidies and other government schemes). 

Since early stage tidal energy deployments often receive a fixed price per MWh, through subsidy schemes described below, most hydrodynamic models used for optimisation of tidal array design produce an average power estimate for the array, $P_{\text{avg}}$, rather than a time varying prediction of the instantaneous power over the whole lifetime. Therefore, the energy generated in each year is assumed to be constant, such that $E_i=E\; \forall i=1,\ldots, L$. In practice, even when evaluating metrics such as NPV and LCOE on an annual level, lunar nodal cycles cause variations in the yearly average yield over an 18.6 year cycle \cite{Thiebot2020InfluenceFrance, Haigh2011GlobalLevels}. Since the earlier years of generation count more towards these metrics, due to discounted cash flow, this could lead to an effective difference in the performance of the array depending on which stage of the tidal cycle it begins generating in.

Assuming a constant power, $P_{\text{avg}}$, year on year, the annual energy generation, $E_i$, can be found by multiplying the number of hours of generation each year, $t_i$, with the average power, to find the gross energy output in MWh; 
\begin{equation} \label{eq:energy}
	\begin{aligned}
    E_i = t_i \times P_{\text{avg}} . 
	\end{aligned}
\end{equation}
$t_i$ can be approximated as the number of hours in a year minus the anticipated number of hours of downtime for maintenance. The MeyGen Phase 1A reported an average availability of 95\% in 2020 (having entered its operational phase in 2018), with a higher availability of 98\% in the summer compared to 90\% in the winter \cite{2020LessonsReport}. 
Increased experience in planning and predicting operations and maintenance could see these downtime windows decrease and availability increase in future tidal projects. 
When used in industry the availability could be modelled to depreciate throughout an array's lifetime, to account for decreases in efficiency as devices age, e.g. due to component wear or bio-fouling. 
This effect has been well documented in the wind industry \cite{Ziegler2018LifetimeUK, Faulstich2011WindDeployment}. However, for simplicity it can be kept constant for initial investigations. This is a reasonable assumption because there is limited information on the rate of degradation of tidal stream arrays, with few years of operational experience to date, and due to the high discount rate currently used in tidal the latter years of the array's generation (where degradation would be highest) contribute significantly less to the NPV and LCOE. 
Finally, the net energy output can be found from the gross energy output by a factor, $C_E$, to account for electrical efficiency losses. 
The net energy output can be multiplied by the electricity tariff, $T_e$ [$\pounds$/MWh], to find the expected revenue of a tidal array in a year. The market price may fluctuate over days and years, however government schemes can often make this value more predictable and stable, the details of these schemes are discussed below in Section \ref{sec:revenue_inputs}. 
In initial economic models it can be assumed that arrays receive a constant price per unit energy, however future models could estimate $P_{\text{avg}}$ over smaller time scales and use this in conjunction with a fluctuating energy price to determine the impact on revenue, cf. related optimisation work in the context of tidal lagoons \cite{Harcourt2019UtilisingValue}. 

According to the simplifications of the economic model made in equations \eqref{eq:CAPEX}, \eqref{eq:OPEX} and \eqref{eq:energy}, the model for LCOE from \eqref{eq:LCOE} can be rewritten as

\begin{equation} \label{eq:LCOE_2}
	\begin{aligned}
    \text{LCOE} =  \frac{\text{CA}_{f}+\text{CA}_{t}\times n_t + \sum_{i=1}^{L} \frac{O_{f}+O_{t}\times n_t}{(1+r)^i}}{\sum_{i=1}^{L} \frac{C_E \times E_i}{(1+r)^i}}   . 
	\end{aligned}
\end{equation}

\section{Estimate of inputs to economic models}\label{sec:input_est}

The following section compiles a series of tidal energy cost estimates and reformats them appropriately for use in a LCOE model such as \eqref{eq:LCOE_2}. The main inputs needed are CAPEX split up into fixed, $\text{CA}_f$, and turbine-dependent, $\text{CA}_t$, components, OPEX split up into fixed, $O_f$, and turbine-dependent, $O_t$, components, lifetime, $L$, and the discount rate, $r$. Data from different sources is discussed below along with the methods used to convert this into the format we need where total CAPEX and OPEX can be broken down into the forms shown in \eqref{eq:CAPEX} and \eqref{eq:OPEX} respectively. Some information from the offshore wind industry is used to guide our estimates and the relationship between them. 

\subsection{Relationship between cost and size of array}

Firstly, information on how CAPEX and OPEX in the wind industry varies in relation to number of turbines is used to justify the relationship assumed in \eqref{eq:CAPEX} and \eqref{eq:OPEX}.  Wind is the more established industry and can therefore be used to help predict the cost reductions that can be realised through economies of volume for tidal. 
CAPEX and OPEX values are commercially sensitive and hard to establish, and while there have been a number of studies which have communicated with many different developers of varying sizes \cite{Smart2018TidalBenefit}, they often anonymise this data by converting to a metric that is independent of array size, such as cost per \mbox{MW} installed capacity. 

\begin{figure}[!t]
\centering
\subfloat[][]{\centering\includegraphics[height=60mm]{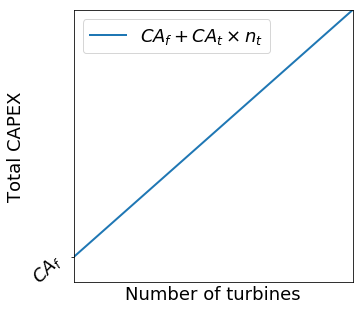}\label{fig:capex_eq_a}} \
\subfloat[][]{\centering\includegraphics[height=60mm]{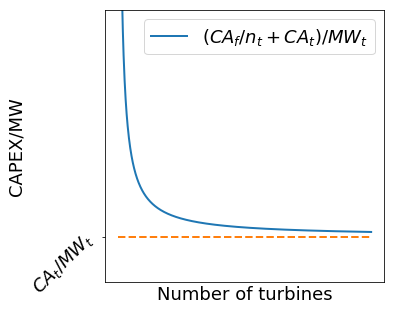}\label{fig:capex_eq_b}} 
\caption{Variation in the total CAPEX of an array and the CAPEX per \mbox{MW} installed capacity as functions of the number of turbines, according to \eqref{eq:CAPEX}}
\label{fig:capex_eq}
\end{figure}

Much of the available information found in this literature review on the costs of tidal turbine arrays is given as a CAPEX/\mbox{MW}, for an array of a certain rated capacity. 
Figure \ref{fig:capex_eq} shows how the total CAPEX increases with number of turbines if the linear relationship given in \eqref{eq:CAPEX} is assumed. This can be used to show the relationship between CAPEX/\mbox{MW} and number of turbines, by dividing equation \eqref{eq:CAPEX} by the total capacity of the array ($n_t$ turbines, each with a rated capacity of $MW_t$) to get
\begin{equation} \label{eq:CAPEX_perMW}
	\begin{aligned}
    \text{CAPEX/MW} = \frac{\text{CA}_{f}+\text{CA}_{t}\times n_t}{n_t \times MW_t}  . 
	\end{aligned}
\end{equation}

Since there are two unknowns in each of \eqref{eq:CAPEX} and \eqref{eq:OPEX}, two simultaneous equations must be used to solve each in order to find $\text{CA}_t$, $\text{CA}_f$, $\text{O}_t$ and $\text{O}_f$. 
In order to convert from the CAPEX/\mbox{MW} to $\text{CA}_f$ and $\text{CA}_t$ values, (estimates for) the CAPEX/\mbox{MW} for arrays of two different sizes is needed, allowing the conversion to total CAPEX, which would give two points on the line in Figure \ref{fig:capex_eq}a) and allow the gradient (i.e. $\text{CA}_t$) and the intercept (i.e. $\text{CA}_f$) to be calculated. Alternatively, if the total CAPEX is only known at one array size then the ratio of the fixed component of CAPEX over the turbine-dependent component of CAPEX, $\text{CA}_f/\text{CA}_t$, must be known or estimated. The ratio can be used to split the total cost into each of its components, such that $\text{CAPEX} =( \frac{\text{CA}_f}{\text{CA}_t} + n_t ) \text{CA}_t$. The same is needed for operational costs to find $O_f$ and $O_t$, however limited OPEX data is available, since very few tidal arrays have been in operation and only for a relatively short period, so sometimes OPEX must be estimated as a percentage of CAPEX. 

\subsubsection{Method 1: Knowing two data points}

If the total CAPEX is known for two values of $n_t$, e.g. $n_{t_1}$, $n_{t_2}$, $\text{CAPEX}(n_{t_1})$ and $\text{CAPEX}(n_{t_2})$, then we can solve the following simultaneous equations:

\begin{equation}\label{eq:CAPEX1}
	\begin{aligned}
    \text{CAPEX}(n_{t_1}) = \text{CA}_{f}+\text{CA}_{t}\times n_{t_1},  
	\end{aligned}
\end{equation}
\begin{equation}
	\begin{aligned}
    \text{CAPEX}(n_{t_2}) = \text{CA}_{f}+\text{CA}_{t}\times n_{t_2}.  
	\end{aligned}
\end{equation}
When subtracting these two equations to solve for $\text{CA}_{t}$, we find
\begin{equation}
	\begin{aligned}
    \text{CAPEX}(n_{t_2})  -\text{CAPEX}(n_{t_1}) = \text{CA}_{t}\times (n_{t_2} -  n_{t_1})
	\end{aligned}
\end{equation}
\begin{equation} \label{eq:cat1}
	\begin{aligned}
    \Rightarrow \text{CA}_{t} = \frac{\text{CAPEX}(n_{t_2})  -\text{CAPEX}(n_{t_1})}{n_{t_2} -  n_{t_1}}
	\end{aligned}
\end{equation}

$\text{CA}_{f}$ can then be found from \eqref{eq:CAPEX1} and the $\text{CA}_{t}$ value just found.

The turbine-dependent and fixed components of the OPEX can be found similarly by evaluating \eqref{eq:OPEX} at two values of $n_t$ and similarly rearranging. 

\subsubsection{Method 2: Knowing the ratio of fixed to turbine-dependent costs}

Alternatively, if the ratio of the fixed costs to the turbine-dependent costs is known, such that
\begin{equation} 
	\begin{aligned}
     \text{CA}_{ft} \coloneq \frac{\text{CA}_{f}}{\text{CA}_{t}},
	\end{aligned}
\end{equation}
then the total CAPEX only needs to be known for one value of $n_t$, e.g. $n_{t_1}$ and $\text{CAPEX}(n_{t_1})$. 
The turbine-dependent CAPEX, $\text{CA}_{t}$, can be found by solving
\begin{equation}
	\begin{aligned}
    \text{CAPEX}(n_{t_1}) = \text{CA}_{ft}\times\text{CA}_{t}+\text{CA}_{t}\times n_{t_1}
	\end{aligned}
\end{equation}
\begin{equation} \label{eq:cat2}
	\begin{aligned}
	\Rightarrow \text{CA}_{t} = \frac{\text{CAPEX}(n_{t_1})}{\text{CA}_{ft}+ n_{t_1}}
	\end{aligned}
\end{equation}
and $\text{CA}_{f}$ is simply found from 
\begin{equation} \label{eq:caf2}
	\begin{aligned}
    \text{CA}_{f} = \text{CA}_{ft}\times \text{CA}_{t}.
	\end{aligned}
\end{equation}
The same relationship can be applied for OPEX.

\subsection{Cost data available in literature}

Using the two methods described above, a literature review of publicly available cost information for tidal stream arrays has been performed. While there are many studies on the economics of tidal energy, not all of them are useful for building an economic model that is flexible to array size. For example, if given the cost of energy as an LCOE, there is rarely enough information to back calculate the raw CAPEX and OPEX values in order to predict how much impact changing the number of turbines would have on the array economics. This would make it very difficult to model the cost reductions that can be found through economies of volume. Instead most studies collated in this work refer to raw CAPEX and OPEX values or CAPEX/MW and OPEX/MW estimates. 

\subsubsection{Higgins and Foley: The evolution of offshore wind power in the United Kingdom} 

\begin{figure}[!t]
\centering
\subfloat[][]{\centering\includegraphics[width=60mm]{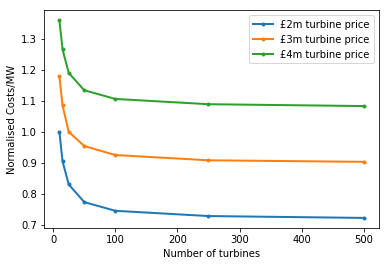}\label{fig:capex_eq_a}} \
\subfloat[][]{\centering\includegraphics[width=60mm]{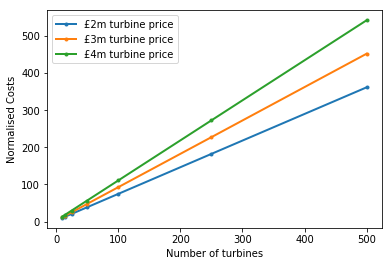}\label{fig:capex_eq_b}} 
\caption{(a) How the normalised capital costs per \mbox{MW} installed capacity varies with the number of turbines from \cite{Higgins2014TheKingdom}.  (b) How this corresponds to a linear relationship between total CAPEX and number of turbines, for wind turbine arrays where each turbine costs £2, 3 or 4 million.}
\label{fig:higgins_capex_234}
\end{figure}

Firstly the assumptions made in \eqref{eq:CAPEX} and \eqref{eq:OPEX}, and demonstrated in Figure \ref{fig:capex_eq}, can be validated by comparing them to actual trends found in the wind industry.  
Higgins and Foley \cite{Higgins2014TheKingdom} performed a review of offshore wind power in the United Kingdom, where they aggregated cost data from multiple wind farms and presented anonymised cost information in the form of a normalised CAPEX/\mbox{MW}.
Higgins and Foley \cite{Higgins2014TheKingdom} investigated how the costs of offshore wind farms increases with cost per device and distance from shore, and decreases with number of turbines.
Figure \ref{fig:higgins_capex_234}(a) shows the relationship they found between number of turbines and normalised CAPEX/\mbox{MW}, and how this varies depending on the price of individual turbines. 
The costs are normalised by the CAPEX/\mbox{MW} of an array consisting of ten turbines each costing £2m, so for example these figures show that if there are ten turbines but they each cost £4m, the CAPEX/\mbox{MW} of the overall array is 36\% higher. 

Figure \ref{fig:higgins_capex_234}(b) shows that when multiplying this by the number of turbines to find normalised total CAPEX values, a linear relationship is obtained, further justifying the relationship defined in \eqref{eq:CAPEX}. 
The exact values of $\text{CA}_f$ and $\text{CA}_t$ cannot be found, since the information was normalised due to commercial sensitives. However, it can be shown that for arrays made of turbines which cost $\pounds$4 million, $\pounds$3 million and $\pounds$2 million, the fixed component of CAPEX would be 2.6, 3.1 and 3.9 times the turbine-dependent component of CAPEX respectively. The cheaper the devices become the more important economies of volume are (since the fixed costs make up a bigger portion of the overall costs). These $\text{CA}_{ft}$ values can be used in method 2 for calculating the cost components.

Higgins and Foley also presented how the CAPEX/\mbox{MW} would change with number of turbines for different distances to shore -- 5km, 15km, 25km, 35km, 45km, 75km and 150km. The southern part of the Pentland Firth, which passes between Stroma Island and the Scottish mainland is only approximately 3km wide. By comparison the narrowest distance between Alderney and Cap de la Hague is 15km wide. Therefore we only consider the $text{CA}_{ft}$ ratios from the CAPEX curves defined from data from turbines 5 to 15km from shore, which have a $\text{CA}_{ft}$ value of 2.3 and 2.7 respectively. As the distance to shore increases the $\text{CA}_{ft}$ ratio increases, since economies of volume become more important to overcome the fixed costs of long cable routing and shipping distances. At 150km to shore $\text{CA}_{ft} = 8.4$, however, this distance is very unlikely for a tidal energy site in the foreseeable future, since the required high energy flows are generally accelerated near islands and other coastal features. Therefore upper and lower bounds of $2.3<\text{CA}_{ft}<3.9$ were used in method 2, for calculating $\text{CA}_f$ and $\text{CA}_t$ separately when given one fixed CAPEX or CAPEX/\mbox{MW} value. There is limited information on OPEX varying with the size of the array so it is assumed that the ratio remains the same. i.e. $\text{O}_{ft} \coloneq \text{CA}_{ft}$.

\subsubsection{Culley et al.: Optimisation of Cable Costs} 

One easy-to-visualise example of how capital costs can be split into a fixed and turbine-dependent component is cable based costs. Culley et al. \cite{Culley2016IntegrationArrays} investigated the optimisation of cable routing and array layout and used estimates for the total cabling costs of between £254.51 per meter with a fixed £5.16 million vessel mobilisation fee and £862.48 per meter with a fixed £6.22 million vessel mobilisation fee. These estimates were based on the costs given for wind energy from Green et al. \cite{Green2007ElectricalPower} in 2007, adjusted for inflation to 2016 levels and with an assumed 51\% rise in costs applied, in line with the 51\% rise in wind energy costs between 2006 and 2010, as outlined by Heptonstall et al. \cite{Heptonstall2012TheFuture}. 
Depending on the weighting of the cable cost in the functional they found optimised configurations of their model of eight turbines with total cable lengths of 9.70km, 9.23km, and 8.71km. This leads to a total cost per metre of $254.51 \times 8.71 \times 1000 = £2.22$ million to $862.48 \times 9.7 \times 1000 = £8.37$ million, or a cost of $£0.28$ to $£1.04$ million per turbine. Therefore for the cable costs alone there is a fixed cost an order of magnitude higher than the per turbine cost. These cost estimates are not included in the calculation of $CA_f$ and $CA_t$ values, because they only cover cable costs, which account for approximately 10\% of the total CAPEX. However, they do help demonstrate why the method of splitting CAPEX into $CA_f$ and $CA_t$ is appropriate.  

\subsubsection{OREC: Tidal Stream and Wave Energy Cost reduction and Industrial Benefit}

A recent cost analysis was performed by the Offshore Renewable Energy Catapult (OREC) based on aggregated data from tidal developers working in the \mbox{kW} to \mbox{MW} scale \cite{Smart2018TidalBenefit}.
They also forecast the costs and job creation of a potential deployment scenario for the UK, where \mbox{1GW} of tidal stream is deployed by 2030 at a rate of \mbox{100MW} per year from 2020. The predicted UK 2030 spend in this scenario was £307 million; with £227 million spent on CAPEX for a \mbox{100MW} array and £80 million on cumulative OPEX for the existing 10 \mbox{100MW} arrays. The study investigated arrays with turbines in the 1--2\mbox{MW} range, so a turbine rating of 1.5\mbox{MW} is assumed in the following calculations. The CAPEX/\mbox{MW} of an \mbox{100MW} array in the OREC report is £2.27 million, which combined with the ratio $2.3<\text{CA}_{ft}$ from \cite{Higgins2014TheKingdom} and using method 2 results in a fixed CAPEX of $£7.6\text{m} \leq \text{CA}_f \leq £12.8\text{m}$ and a turbine-dependent CAPEX of $£3.26\text{m} \leq \text{CA}_t \leq £3.34\text{m}$. The OPEX/\mbox{MW} of a \mbox{100MW} array is £0.08 million/year, which can be similarly split into $£0.27\text{m} \leq O_f \leq £0.45\text{m}$ and $£0.113\text{m} \leq O_t \leq £0.116\text{m}$.

\subsubsection{IEA Technology: International LCOE for ocean energy technology}

An IEA Technology report estimated costs for different ocean technologies at different stages of their development \cite{2015InternationalTechnology}. 
Looking at the first commercial scale projects with a capacity of \mbox{3--90MW} and assuming that the maximum project capacity corresponds to the minimum CAPEX/\mbox{MW}, and vice-versa, their minimum and maximum scenarios correspond to a small array with two 1.5\mbox{MW} turbines and a larger array of 60 turbines with a total CAPEX of \$16.8m and  \$297m, respectively. 
These two points can be converted to GBP (assuming a rate of \$1 = £0.79) and used in method 1 to find a fixed CAPEX of $\text{CA}_f = £5.6m$ and a turbine-dependent CAPEX of $\text{CA}_t = £3.8m$.

Similarly, IEA Technology found a small (3\mbox{MW}) commercial array has an OPEX/year of \$1.2m and the large array (90\mbox{MW}) is \$8.1m/year. From this, method 1 can be use to calculate the fixed OPEX per year as $O_f =$ £0.76m/year and the turbine-dependent OPEX per year as $O_t =$ £0.094m/year. 

\subsubsection{Orbital Marine Power Limited: Technology Update}

Orbital Marine Power Limited (formerly known as Scotrenewables Tidal Power) issued a cost analysis for a 10MW array, with their cost assumptions and methodology approved by the Carbon trust. It predicted a base CAPEX/MW of £2.60m with a pessimistic prediction of £2.80m and an optimistic one of £2.50m \cite{MarkHamilton-CTO2012TechnologyUpdate}. Furthermore they had a base OPEX/\mbox{MW}/year prediction of £0.1982m with a pessimistic value of £0.235m and an optimistic value £0.16m. Since this range of CAPEX and OPEX predictions were only for \mbox{10MW} arrays, they can be combined via method 2 with the $2.3<\text{CA}_{ft}<3.9$ ratio from Higgins and Foley \cite{Higgins2014TheKingdom}, which defines the shape of the cost reductions with economies of volume. This results in six estimates of each of $\text{CA}_f$, $\text{CA}_t$, $O_f$ and $O_t$, which are summarised in Table \ref{tab:orbital}. Taking the extremes for each value, the fixed CAPEX may lie in the range $£6.3\text{m}< \text{CA}_f <£10.4\text{m}$, turbine-dependent CAPEX may be $£2.4m< \text{CA}_t <£3.1m$, fixed OPEX per year may be $£0.41\text{m}< O_f <£0.87\text{m}$, and turbine-dependent OPEX per year may be $£0.15\text{m}< O_t <£0.26\text{m}$.

\begin{table}[!t]
\caption{The CAPEX and OPEX components found by combining the optimistic (Opt), typical (Typ) and pessimistic (pes) estimates of Orbital Marine Power Ltd \cite{MarkHamilton-CTO2012TechnologyUpdate}, with the upper and lower $\text{CA}_{ft}$ limits found in \cite{Higgins2014TheKingdom}.}\label{tab:orbital}
\begin{tabular}{l|lll|lll}\hline
 & \multicolumn{3}{c}{$\text{CA}_{ft} = 2.3$} & \multicolumn{3}{c}{$\text{CA}_{ft} = 3.9$}\\
 & Opt & Typ & Pes & Opt & Type & Pes \\\hline
$\text{CA}_f$ (£m) & 6.4 & 6.7 & 7.2 & 9.2 & 9.6 & 10.3 \\
$\text{CA}_t$ (£m) & 2.4 & 2.5 & 2.6 & 2.8 & 2.9 & 3.1 \\
$O_f$(£m/year) & 0.41 & 0.49 & 0.60 & 0.59 & 0.71 & 0.87 \\
$O_t$(£m/year) & 0.15 & 0.18 & 0.22 & 0.18 & 0.21 & 0.26 \\\hline
\end{tabular}
\end{table} 

\subsubsection{Black \& Veatch: Cost of and financial support for wave, tidal stream and tidal range generation in the UK}

An executive summary by Ernst \& Young and Black \& Veatch \cite{2010CostBackground} investigated the relative costs of wave and tidal energy in the UK, at different stages of deployment. They provided cost estimates for tidal arrays in both deep ($\leq 40m$) and shallow ($> 40m$) water for a \mbox{10MW} tidal stream array at different stages of developer experience; for a developer's first \mbox{10MW} demonstration project and for a \mbox{10MW} commercial project where the developer has already deployed over \mbox{50MW} in the past. Since these economic models are to be used to optimise large-scale tidal arrays, commercial figures are used, and because the likely locations of the UK's first commercial tidal arrays, e.g. in the Alderney Race and Pentland Firth, are approximately 30--40m in depth, the shallow water values are used in this work. 

Black \& Veatch estimated that the typical CAPEX/\mbox{MW} for a \mbox{10MW} tidal stream array is £3.2m/\mbox{MW}, with a lower and upper bound of £2.7m/\mbox{MW} and £3.9m/\mbox{MW}, respectively. By comparison they estimated that a developer's first \mbox{10MW} demonstrator array would have a CAPEX/\mbox{MW} of £4.3m/\mbox{MW}, showing the significant impact that learning-by-doing can have on the cost of tidal stream in the short term. A commercial array in deep water rather than shallow saw only a small increase in CAPEX/\mbox{MW} to £3.3m/\mbox{MW}, which was attributed to a different deployment method, structure foundations or mooring possibly being needed in greater depths, but otherwise the shallow and deep technologies may be largely the same, and can therefore benefit from learning from each other. 

Similarly, Black \& Veatch estimated that the typical OPEX/\mbox{MW} for a \mbox{10MW} tidal stream array is £150k/\mbox{MW}, with a lower and upper bound of £120k/\mbox{MW} and £190k/\mbox{MW}, respectively. Demonstrator arrays are substantially more expensive with an OPEX/\mbox{MW} of £310k/\mbox{MW}. Deeper projects were predicted to have a lower OPEX/\mbox{MW} of £120k/\mbox{MW}. Black \& Veatch noted that all costs were expected to decline due to the anticipated future global deployments and corresponding impact on learning in the industry. 

These typical and lower/upper bounds can be converted into their separate fixed and turbine-dependent components using method 2 and the $2.3<\text{CA}_{ft}<3.9$ ratio from Higgins and Foley \cite{Higgins2014TheKingdom}. The three values given for each of CAPEX/\mbox{MW} and OPEX/\mbox{MW} by Black \& Veatch, along with the upper and lower values for $\text{CA}_{ft}$ result in six estimates for each parameter, which are summarised in Table \ref{tab:BandV}. The maximum and minimum values in this table provide the parameter ranges that can be estimated using the Black \& Veatch cost information for commercial arrays. The fixed CAPEX may lie in the range $£6.9\text{m}< \text{CA}_f <£14.5\text{m}$, turbine-dependent CAPEX may be $£2.5\text{m}< \text{CA}_t <£4.4\text{m}$, fixed OPEX per year may be $£0.31\text{m}< O_f <£0.70\text{m}$, and turbine-dependent OPEX per year may be $£0.11\text{m}< O_t <£0.21\text{m}$.

\begin{table}[!t]
\caption{The CAPEX and OPEX components found by combining the optimistic (Opt), typical (Typ) and pessimistic (Pes) estimates of Black \& Veatch \cite{2010CostBackground}, with the upper and lower $\text{CA}_{ft}$ limits found in \cite{Higgins2014TheKingdom}.}\label{tab:BandV}
\begin{tabular}{l|lll|lll}\hline
 & \multicolumn{3}{c}{$\text{CA}_{ft} = 2.3$} & \multicolumn{3}{c}{$\text{CA}_{ft} = 3.9$}\\
 & Opt & Typ & Pes & Opt & Type & Pes \\\hline
$\text{CA}_f$ (£m) & 6.9 & 8.2 & 10.0 & 10.0 & 11.8 & 14.5 \\
$\text{CA}_t$ (£m) & 3.0 & 3.6 & 4.4 & 2.5 & 3.0 & 3.7 \\
$O_f$(£m/year) & 0.31 & 0.38 & 0.49 & 0.44 & 0.55 & 0.70 \\
$O_t$(£m/year) & 0.13 & 0.17 & 0.21 & 0.11 & 0.14 & 0.18 \\\hline
\end{tabular}
\end{table} 

\subsubsection{Summary of $\text{CA}_t$, $\text{CA}_f$, $O_t$ and $O_f$ estimates}

This literature review sought to collate CAPEX/\mbox{MW} and OPEX/\mbox{MW} estimates from a range of sources and convert them to a format more suitable for modelling the impact of economies of volume, to enable the comparison of arrays of different sizes.
Table \ref{tab:econvals} summarises the upper and lower bounds and the average of the different estimates for $\text{CA}_t$, $\text{CA}_f$, $O_t$ and $O_f$ found in each of the sources reviewed above. 

\begin{table}[!t]
\caption{A summary of the typical (Typ), optimistic (Opt) and pessimistic (Pes) estimates of $\text{CA}_t$, $\text{CA}_f$, $O_t$ and $O_f$ from IEA Technology \cite{2015InternationalTechnology}, Orbital Marine Power Limited \cite{MarkHamilton-CTO2012TechnologyUpdate}, Black \& Veatch \cite{2010CostBackground} and OREC 2018 report \cite{Smart2018TidalBenefit}.}\label{tab:cafcat_sources_summary2}
\begin{tabular}{c|cc|c|cc|cc}\hline
Source: & \multicolumn{2}{c}{OREC} & \multicolumn{1}{c}{IEA}  & \multicolumn{2}{c}{Orbital Marine Power Ltd} & \multicolumn{2}{c}{Black \& Veatch} \\
Type: & Opt & Pes & Typ & Opt & Pes & Opt & Pes \\\hline
$\text{CA}_f$ (£m) & 7.6 & 12.8 & 5.6 & 6.3 & 10.4 & 6.9 & 14.5 \\
$\text{CA}_t$ (£m) & 3.26 & 3.34 & 3.8 & 2.4 & 3.1 & 2.5 & 4.4 \\
$O_f$(£m/year) & 0.27 & 0.45 & 0.76 & 0.41 &  0.87 & 0.31 & 0.70  \\
$O_t$(£m/year) & 0.113 & 0.116 & 0.094 & 0.15 &  0.26 & 0.11 & 0.21 \\\hline
\end{tabular}
\end{table} 

\subsubsection{Discount rate}

The discount rate, $r$, is a key parameter when evaluating the economic success of a project over its lifetime as it determines the present value of future cash flows. It is the interest rate used in discounted cash flow analysis. It reflects that future cash flow has less value than current cash flow to investors due to opportunity costs and risks. In general higher rates are applied to less developed technologies, to account for the greater risks and uncertainty associated with the novel design and speculative cost estimation. Determining an appropriate discount rate is important for any economic metrics such as net present value, LCOE and payback periods. 

Ouyang and Lin \cite{Ouyang2014LevelizedChina} investigated the LCOE of different renewable energy technologies in China and the appropriate subsidies and policies to support them. They found the discount rates required for different forms of renewables varied through 5, 8 and 10\%, with the higher rates being needed for novel renewable energy sources, such as tidal, which are seen as more risky compared to more established forms of renewables, such as onshore wind or solar PV. In general it can be shown that a transition to lower discount rates will help renewable energy become more cost competitive with fossil fuels. Khatib \cite{Khatib2010ReviewEdition} performed a review of generation costs in OECD countries and found that as the discount rates fall from 10\% to 5\% more capital intensive forms of energy, such as nuclear, wind and tidal, become more cost competitive compared to coal and gas. Both papers used a common discount rate across the whole energy market due to limited data to distinguish the relative risks for different technologies. Common discount rates have been criticised for not reflecting these risks appropriately, 
and more recent studies have suggested specific values for the tidal other ocean energy industries, which are on the higher end of the spectrum due to their novelty and perceived risk.

A report on the cost of ocean energy, by SI Ocean for the European Commission \cite{2013OceanM}, used a discount rate of 12\% for both wave and tidal, but also investigated how changes in the rate chosen could have a significant impact on the LCOE. Reducing $r$ from 12\% to 6\% results in a decrease in their predicted LCOE of demonstrator arrays, from 32.0c/kWh (£288/MWh) to 23.1c/kWh (£208/MWh). 

The OREC 2018 report estimated a discount rate at \mbox{10MW} cumulative tidal capacity of 10\%, but their models predicted this would fall to 8.4\% by \mbox{100MW}, 8.0\% by \mbox{200MW} and 7.1\% by \mbox{1GW} global installed capacity, reflecting the potential that learning has to reduce the cost of capital. 
They also demonstrated that each percentage point reduction in discount rate leads to a significant LCOE reduction (of approximately 6\%), for example the discount rate changing from 7.1\% to 6.1\% would reduce their \mbox{1GW} cummulative capacity LCOE estimate from £91 per MWh to £80 per MWh\cite{Smart2018TidalBenefit}. This is consistent with SI Oceans's results, which fell by 6.4\% on average per percentage point reduction in discount rate. Falls in $r$ could be achieved as the industry matures and investing in tidal energy is seen as less of a risk. 

The Carbon Trust predicted that the first commercial marine energy schemes would have a discount rate of around 15\%, which could fall to 8\% as the technology matures \cite{2006CostMethodology}. 
Vazquez and Iglesias \cite{Vazquez2016CapitalApproach,Vazquez2015LCOEEnergy} used a discount rate of 10\% in their Levelised CAPEX of Energy tool. In another paper they argued that tidal stream energy projects have greater technological risks, due to their novelty compared to conventional types of energy generation, and high capital costs which results in a conservative discount rate being needed of between 10\% and 12\% \cite{Vazquez2015DeviceEnergy}.

Allan et al. \cite{Allan2011LevelisedCertificates} used a high and low discount rate of 15\% and 6\% respectively, with 10\% used as the central value when finding the LCOE of wave and tidal, which is often used as a common rate across multiple different technologies \cite{IEANEA2015Projected2015,2016ELECTRICITYCOSTS}. Allan et al. \cite{Allan2011LevelisedCertificates} noted that higher discount rates adversely affect technologies with longer lifetimes and high CAPEX as a proportion of the LCOE. Both of these factors apply to tidal, so it is anticipated to be highly sensitive to variation in the discount rate. 

Culley et al. \cite{Culley2016IntegrationArrays} also used a 10\% discount rate for their study into cost modelling and micro-siting of tidal stream, and found that this amounted to a 35\% reduction in income when compounded over an assumed 20 year array lifetime. Coles et al. \cite{Coles2019MechanismsEnergy} used a discount rate of 12\% when investigating mechanisms for reducing the LCOE of tidal. Dalton et al. \cite{Dalton2015EconomicPerspectives} suggested a discount rate of 8\% to 15\% is typical in UK ocean energy, however they noted that more in-depth economic studies could use multiple rates within one project to reflect the different risks of individual cash flows. Klaus et al. \cite{Klaus2020FinancialStudy} assumed a financial discount rate of 10\% but also tested the impact of varying it over the range 5--15\%. They emphasised the importance of thorough investigation into the choice of discount rate, by demonstrating the the results of LCOE comparisons between technologies can be inverted as the discount rate is varied through its conventional range.  

This present study therefore recommends using a discount rate of 5 to 15\% with a typical value of 10\%. However, due to the high sensitivity of LCOE to the discount rate, sensitivity analysis of the final LCOE prediction to the choice of $r$ should be performed. 

\subsubsection{Lifetime of array}

The lifetime of the array is the number of years that the tidal project is planned to operate for. It is normally determined by contracts and insurance based on the assumed time an array can perform well before there is too much degradation due to environmental conditions. Vazquez and Iglesias \cite{Vazquez2016CapitalApproach, Vazquez2015DeviceEnergy, Vazquez2015LCOEEnergy} used an expected lifetime of installation of 20 years in their Levelised CAPEX of Energy tool. Likewise Dalton et al. \cite{Dalton2015EconomicPerspectives} and the report by SI Ocean\cite{2013OceanM} assumed a lifetime of 20 years. The Department of Business, Energy and Industry Strategy (BEIS) assumed an operating period of 22 years for tidal stream in their 2016 generation costs report\cite{2016ELECTRICITYCOSTS}. 

The world's largest currently operating, the MeyGen \mbox{6MW} array in the Pentland Firth, announced in 2018 it had entered into its 25 year operations phase \cite{MeyGenEnergy,2020LessonsReport}. Similarly Johnstone et al. \cite{Johnstone2013ATechnology} assumed a lifetime of 25 years in their techno-economic analysis of tidal energy, and as did Coles et al. \cite{Coles2019MechanismsEnergy}. As the technology becomes more tested and proven lifetimes are likely to increase even further. Therefore a typical value of 25 years with an upper and lower bound of 30 and 20 years respectively was used in this study. 

Tidal energy is currently perceived as a relatively risky investment, and therefore has a high discount rate, this makes the NPV and LCOE less sensitive to the choice of array lifetime, because both revenue and costs many years into the future are heavily discounted. At a discount rate of 10\% cashflows 20, 25 and 30 years into the future have a present value reduced by 85\%, 91\% and 94\% respectively, so adding additional years to the project lifetime does not contribute much to the overall LCOE. If the discount rate falls to 5\% then the present value reductions fall to 62\% 70\% and 77\% respectively. 

\subsubsection{Summary of all cost model inputs}

The maximum, minimum and mean values of the cost parameters and other economic inputs (discount rate and lifetime of an array) across all the sources discussed above are summarised in Table \ref{tab:econvals}, as the Pessimistic, Optimistic and Typical values. It should be noted that the cost information summarised in this review is based on the limited publicly available information at the time of writing and that there is a significant difference between the optimistic and pessimistic values due to the high degree of uncertainty in this relatively new industry. These estimates are suitable for academic modelling and optimisation of tidal stream arrays, but in practice tidal developers should use their internal cost information for more accurate economic array design.

\begin{table}[!t]
\caption{Estimates for the parameters used in the economic models, and the amount they vary.}\label{tab:econvals}
\begin{tabular}{llllll}\hline
Symbol & Description & \multicolumn{3}{c}{Value range} & Units\\\hline
 & & Optimistic & Typical & Pessimistic  & \\\hline
$\text{CA}_f$ & Fixed CAPEX & 5.6 & 9.2 & 14.4 & £m \\
$\text{CA}_t$ & Turbine dependent CAPEX & 2.4 & 3.3 & 4.4 & £m per turbine \\
$O_f$ & Fixed OPEX & 0.27 & 0.32 & 0.87 & £m per year \\
$O_t$ & Turbine dependent OPEX & 0.094 & 0.15 & 0.26 & £m per year per turbine \\
$r$ & Discount rate & 0.05 & 0.10 & 0.15 & N/A \\
$L$ & Lifetime of an array & 30 & 25 & 20 & years \\\hline
\end{tabular}
\end{table}

\subsection{Revenue inputs}\label{sec:revenue_inputs}

In all of the financial models above, the income must be estimated, to be able to analyse the cost-benefit balance of different array designs. Assuming that the main source of income for an array operator is the from selling the energy generated, then the revenue, in year $i$, can be approximated from

\begin{equation} \label{eq:revenue}
	\begin{aligned}
    \text{Revenue}_i = P_{\text{avg}} \times t_i \times T_e . 
	\end{aligned}
\end{equation}

Where $P_{\text{avg}}$ is the average power generated, predicted using a hydrodynamic model of the array, and either used as an average over the whole lifetime of the array, or calculated over smaller time scales, for example to take into account variation due to the lunar cycle. $t_i$ is the number of hours of generation each year, which can be multiplied by the average power, to find the gross energy output in MWh. Below is a discussion on factors that affect the ratio of operational hours to hours downtime. $T_e$ [$\pounds$/MWh] is the electricity tariff, i.e. the price per MWh the electricity generated is sold at. This price is greatly dependent on the subsidies and other schemes available to developers, and the schemes available in the UK are discussed in greater detail below.

\subsubsection{Strike price of energy}

Traditionally, large scale electricity generators sell their energy to suppliers at a time varying market price. In the UK, the wholesale electricity market price (or `reference price') hovers between £40--£50/MWh. This market price changes throughout the day and the year in response to the balance of supply and demand. 

Renewable energy sources often need subsidies and schemes for additional income support for a number of reasons. Firstly, many forms of renewables, especially the tidal industry, have a relatively high LCOE due to their relative infancy. Governments often seek to subsidise the expensive new forms of energy by paying higher than the market rate. This helps decarbonise the electricity grid and support in the early stages may help new technologies become cost competitive by the time they are deployed on a larger scale. Secondly, renewable energy technologies are often non-dispatchable, in that the production cannot be scaled up and down on demand easily. This means they cannot react to falls and rises in the market prices, which can lead to worsening problems as renewables account for increasingly high portions of the energy generation mix. The wind energy industry has already seen that during times where the wind is high at one farm often correlates with high winds at other farms in the surrounding regions, and therefore the price falls and operators get a low return per unit during their periods of highest production. In the US this has led to a rise in the frequency of negative prices in areas with high levels of wind and solar and which have transmission constraints \cite{Mills2019Impact2017}. 
To address this problem governments can offer schemes that guarantee a fixed electricity tariff. These tariffs are not necessarily higher that the market price, but the guarantee reduces the uncertainty the investors and can result in lower discount rates and protect the income of operators during times where supply is high. This is less likely to be a problem for tidal energy because it accounts for a far lower proportion of total energy generation than wind energy and it is possible to take advantage of phase differences across the country to smooth out the daily power production\cite{Neill2014OptimalFunction}. Improvements to the grid, such as increased storage, may reduce the need for these kind of subsidies in the future.

Power Purchase Agreements (PPAs) are contracts between the generators and the suppliers that outline how much generators will be paid for the power exported over the duration of the contract. PPAs enable generators to earn payments for the energy they export to the grid. These agreements can be long or short term, so could be negotiated to last the full lifetime of a tidal array. PPAs ca be negotiated to be market-varying price or fixed price, the latter being preferable for a predictable income for tidal projects. Similarly to ROCs, PPAs help remove exposure to price volatility and make cost planning more predictable. The advantages of PPAs from a supplier's perspective include fixed prices to reduce exposure to makret fluctuations and guarantees of renewable origin, to help meet the business's sustainability goals. MeyGen Phase 1A in Pentland Firth signed a 10-year PPA with SmartestEnergy, to guarantee revenue on the power generated up to 2025. DeltaStream in Wales signed a PPA with EDF for their first tidal turbine in 2014. 

The Renewables Obligation (RO) scheme incentivised large-scale ($\geq 5$\textbf{MW} capacity) renewable electricity in the UK by requiring UK electricity suppliers to source a specified proportion of the electricity they provide to customers from eligible renewable sources. The scheme closed to new generators in March 2017, but electricity sources already accredited under ROs will  receive their 20 year lifetime support until the final close of the scheme in 2037. Under this scheme, Renewable Obligation Certificates (ROCs) are issued to accredited renewable energy generators for the eligible electricity they generate. Generators can sell their ROCs to suppliers through the e-ROC auction, receiving a premium ontop of the wholesale electricity price. Suppliers present the ROCs bought from generators to Ofgem, and if they do not have enough ROCs to cover their obligation they must make a payment into the buy-out fund at a fixed price per MWh. Ofgem then redistributes the buy out funds proportionately to the suppliers who presented sufficient ROCs. When first introduced in 2002, one ROC is issued to the generators per MWh generated, to emphasise competition between technologies. However this favoured more established technologies, so in 2006 it was reformed to have different banding levels for different types of technology. Tidal stream projects accredited before the scheme closed in 2017 were issued 5 ROCs (the highest banding level), subject to a \textbf{30MW} cap at each generating station. Power generated above the \textbf{30MW} cap receives 2 ROCs. PPAs with a particular energy supplier can also guarantee the purchase of the ROCs issued to that generator over a specified period.
The MeyGen Phase 1A was awarded Ofgem accreditation as a 5 ROC project in 2017, meaning that their revenue was comprised of power generation sales (as agreed by a PPA) and a buy-out price of £44.77/MWh from each of the ROCs (which has since risen to £50.80/MWh in 2021/22), which would bring the total $T_e$ is circa £250 to £300/MWh, depending on the price agreed in the PPA.

As the ROC scheme was phased out in 2017 it was replaced with the Contracts For Difference (CfD) scheme, for guaranteeing low-carbon electricity generators a long-term energy price. A fixed ``strike price'' for each unit of energy produced is guaranteed, leading to a promise of steady returns for investors. During the contract period, typically 15 years, a low-carbon energy project is paid the difference between the market price and the strike price (or they pay back the difference, if the market price rises above the strike price).
In the UK's most recent round (the third allocation in 2019), CfDs were allocated as low as £39.65/MWh, mostly for offshore wind. In its current state of deployment, the tidal industry cannot compete with these low prices. Alternative subsidy schemes, which allow developers of new forms of green energy greater funding until their costs can fall due to learning and economies of volume, are being investigated. 

A recent Marine Energy Council report \cite{2019UKIndustry} proposed a number of different subsidy schemes to provide a route to market for the tidal energy industry and other innovative clean energy technology types, such as wave and Advance Combustion Technologies.
An Innovation Power Purchase Agreement (IPPA) could support small-scale (up to \mbox{5MW}) novel projects by starting at a guaranteed price, far above the market rate, at prices starting at £290/MWh and falling by 15\% for every 30MW of deployment (where each individual technology type can only make use of 5MW out of every 30MW price band) down to £150/MWh by 120MW of net deployment. The idea would help enable each novel technology to demonstrate their performance and cost reduction potential, without having to compete against far more established technologies. 

Marine Energy Council also proposed an Innovation Contract for Difference (iCfD) to allow for a new pot within the government's existing CfD framework for new technologies \cite{2019UKIndustry}. This would act as a bridging mechanism to allow projects of larger than \mbox{5MW} but less than \mbox{100MW} and exploit economies of volume until they can compete in open CfD rounds or other PPAs which could enable a higher revenue than the UK market price. It could be limited to projects of up to \mbox{100MW} and could see the costs fall by 7\% per \mbox{100MW} from £150/MWh to £90/MWh, based on the predictions of the OREC 2018 report \cite{Smart2018TidalBenefit}. Marine Energy council anticipated that these proposed schemes would support tidal energy to reach target costs of less than £100MWh after 1GW of deployment

These wide ranges of prices represent the highs and lows of what may be achievable through subsidy for tidal energy in different stages of global deployment. This study proposes a typical strike price of £150/MWh, chosen to represent the gap bridged between small scale demonstrator arrays and larger ones which can use iCfDs, with a lower bound of £40/MWh and an upper bound of £290/MWh to represent the market prices or open CfDs and the propose iPPA starting prices respectively. However it should be noted that there is a great deal of uncertainty in these prices and they will depend greatly on the state of the marine energy industry and government decisions at the time. Using LCOE for evaluating the economics of tidal arrays removes the need for assuming the strike price of energy in the calculations, so it can be a suitable metric until more information about the subsidy levels available are known. 

\subsubsection{Downtime and degradation}\label{sec:downtume}

Availability is a measure of the time that an array is available for operation. Availability can be calculated from the number of hours downtime over a given period, such that
\begin{equation} \label{eq:availability}
	\begin{aligned}
    \text{Availability} = 1 - \frac{\text{hours downtime}}{\text{total hours}}  ,
	\end{aligned}
\end{equation}
therefore the number of hours of generation in year $i$, as used in the LCOE expression \eqref{eq:LCOE}, can be predicted as $t_i = 365\times24$ multiplied by the anticipated availability in year $i$. In the wind industry it has been found that turbines often have lower availabilities during high-wind periods, where the production and loads are higher so faults are more likely to occur. To account for this effect, availability is sometimes calculated in two ways -- as a time-based availability given in \eqref{eq:availability} or as a production-based availability, found as the percentage of actual energy produced over energy expected. The former is easier to calculate but the latter is a better representation of the energy lost. Downtime in high winds often results in the percentage energy lost being higher than the percentage of hours lost, with one study finding that a 3\% non-availability in time resulted in an 11\% reduction in energy generated in the Irish wind farm investigated \cite{Conroy2011WindIreland}, whereas DNV-GL found that the two metric differed by up to 2\%\cite{2017DefinitionsIndustry}. The production-based availability can be improved by scheduling maintenance, where possible, during periods of low resource. 

Hours of downtime, or non-availability, can have several causes; turbine availability can include scheduled or unscheduled maintenance or faults in the turbines causing periods of non-operation, grid availability can include periods of time where the grid is unable to accept electricity due to lack of capacity or grid failure, or balance of plant (BoP) availability, where electricity generated at the turbines is lost due to failure of supporting components and auxiliary systems \cite{2017DefinitionsIndustry}.  

Turbine suppliers will often guarantee a minimum turbine availability rate when they sell their turbines to operators and if turbine failures cause the availability to drop below that value the suppliers will pay compensation to the operators. This contractual availability is negotiated during the turbine supply agreement (TSA). In wind energy a typical value of 97\% is used as the industry standard \cite{Conroy2011WindIreland}. There is limited operational data to form conclusions about the typical time-based versus production-based availability in tidal, but the MeyGen Phase 1A guaranteed a contractual availability of 95\%, and anticipated that the turbines would exceed their target performance in practice\cite{2017LessonsPhase,2020LessonsReport}.

When more data becomes available from operational tidal stream arrays, it may become possible to model the downtime as a function of time. There are likely to be cyclic patterns to the number of hours downtime needed a year for scheduled maintenance, as some operations may need to occur on a five year cycle for example. There is also evidence from the wind industry that failure rates vary greatly depending on the year in the operating lifetime \cite{Ziegler2018LifetimeUK}. Faulstich et al. \cite{Faulstich2011WindDeployment} demonstrated that wind farms often follow a ``bathtub curve'' where there is a high failure rate in the early life due to teething problems or `infant mortality', a period where the failure rate is approximately constant and low, with just intrinsic random failures occurring and a wear-out period near the end of an array's lifetime, where damage accumulates and the failure rate increases. This degradation could be due to increased component wear or bio-fouling impacts. 

\section{Conclusions and summary}\label{sec:conclusion}

There are a great number of metrics that can be used to evaluate the performance of tidal energy arrays. These include power alone (which results with array designs with too many turbines if the number is not pre-specified), purely economic metrics such as break-even power analysis, NPV, LCOE, IRR and PP. Some studies have expanded upon performance metrics further to include the trade-off between economic performance and environmental impacts of the arrays, however this requires decisions on the relative weightings each criteria.
This study proposes that many of these economic metrics can be estimated for large scale arrays by assuming a linear relationship between CAPEX and OPEX and the number of turbines, an therefore splits each of these expenditures into their fixed and turbine-dependent components, $CA_f$, $CA_t$, $O_f$ and $O_t$ respectively.
The data collection study identifies a realistic range for each of the parameters needed in a simple economic model of a tidal stream array, summarised in Table \ref{tab:econvals}. Additionally, the range of past, present and proposed subsidy schemes in the UK were examined, to identify possible upper and lower bounds on strike prices, however it is noted that the actual value will be highly dependent on the state of the industry and levels if government support at the time. It should be noted that there is a great deal of uncertainty in each of the economic input estimates and they should be used for the purpose of providing a reasonable range for academic studies and may not reflect the real economic performance of a potential tidal site. These estimates are useful in the absence of real financial data from developers, which is often highly commercially sensitive, and can be used for proof of concept when demonstrating new techniques for optimising tidal array designs. In practice, array design studies should be repeated with a developers internally-validated financial models.    

\section{Acknowledgements}

ZLG acknowledges the support of Engineering and Physical Sciences Research Council (EPSRC) Centre for Doctoral Training in the Mathematics of Planet Earth [grant number EP/L016613/1].

DSC acknowledges the financial support of the Tidal Stream Industry Energiser project (TIGER), co-financed by the European Regional Development Fund through the Interreg France (Channel) England Programme.

MDP would like to acknowledge the support of the UK's Engineering and Physical Sciences Research Council (EPSRC) [grant numbers EP/M011054/1, EP/L000407/1 and EP/R029423/1].

\bibliographystyle{IEEEtran}
\bibliography{references,sample}

\begin{thebibliography}{10}
\providecommand{\url}[1]{#1}
\csname url@samestyle\endcsname
\providecommand{\newblock}{\relax}
\providecommand{\bibinfo}[2]{#2}
\providecommand{\BIBentrySTDinterwordspacing}{\spaceskip=0pt\relax}
\providecommand{\BIBentryALTinterwordstretchfactor}{4}
\providecommand{\BIBentryALTinterwordspacing}{\spaceskip=\fontdimen2\font plus
\BIBentryALTinterwordstretchfactor\fontdimen3\font minus
  \fontdimen4\font\relax}
\providecommand{\BIBforeignlanguage}[2]{{%
\expandafter\ifx\csname l@#1\endcsname\relax
\typeout{** WARNING: IEEEtran.bst: No hyphenation pattern has been}%
\typeout{** loaded for the language `#1'. Using the pattern for}%
\typeout{** the default language instead.}%
\else
\language=\csname l@#1\endcsname
\fi
#2}}
\providecommand{\BIBdecl}{\relax}
\BIBdecl

\bibitem{2019ContractsResults}
\BIBentryALTinterwordspacing
``{Contracts for Difference Allocation Round 3 Results},'' Tech. Rep., 2019.
  [Online]. Available:
  \url{https://www.gov.uk/government/publications/contracts-for-difference-cfd-allocation-round-3-results}
\BIBentrySTDinterwordspacing

\bibitem{Funke2016DesignApproach}
\BIBentryALTinterwordspacing
S.~W. Funke, S.~C. Kramer, and M.~D. Piggott, ``{Design optimisation and
  resource assessment for tidal-stream renewable energy farms using a new
  continuous turbine approach},'' \emph{Renewable Energy}, vol.~99, pp.
  1046--1061, 2016. [Online]. Available: \url{doi:10.1016/j.renene.2016.07.039}
\BIBentrySTDinterwordspacing

\bibitem{Funke2014TidalApproach}
\BIBentryALTinterwordspacing
S.~W. Funke, P.~E. Farrell, and M.~D. Piggott, ``{Tidal turbine array
  optimisation using the adjoint approach},'' \emph{Renewable Energy}, vol.~63,
  pp. 658--673, 3 2014. [Online]. Available:
  \url{doi:10.1016/j.renene.2013.09.031}
\BIBentrySTDinterwordspacing

\bibitem{Culley2016IntegrationArrays}
\BIBentryALTinterwordspacing
D.~M. Culley, S.~W. Funke, S.~C. Kramer, and M.~D. Piggott, ``{Integration of
  cost modelling within the micro-siting design optimisation of tidal turbine
  arrays},'' \emph{Renewable Energy}, vol.~85, pp. 215--227, 2016. [Online].
  Available: \url{doi:10.1016/j.renene.2015.06.013}
\BIBentrySTDinterwordspacing

\bibitem{duFeu2017TheProblem}
\BIBentryALTinterwordspacing
R.~J. du~Feu, S.~W. Funke, S.~C. Kramer, D.~M. Culley, J.~Hill, B.~S. Halpern,
  and M.~D. Piggott, ``{The trade-off between tidal-turbine array yield and
  impact on flow: A multi-objective optimisation problem},'' \emph{Renewable
  Energy}, vol. 114, pp. 1247--1257, 12 2017. [Online]. Available:
  \url{doi:10.1016/j.renene.2017.07.081}
\BIBentrySTDinterwordspacing

\bibitem{Goss2018CompetitionRace}
Z.~L. Goss, M.~D. Piggott, S.~C. Kramer, A.~Avdis, A.~Angeloudis, and C.~J.
  Cotter, ``{Competition effects between nearby tidal turbine arrays-optimal
  design for Alderney Race},'' in \emph{3rd International Conference on
  Renewable Energies Offshore}, Lisbon, 2018, pp. 255--262.

\bibitem{Divett2013OptimizationMesh}
\BIBentryALTinterwordspacing
T.~Divett, R.~Vennell, and C.~Stevens, ``{Optimization of multiple turbine
  arrays in a channel with tidally reversing flow by numerical modelling with
  adaptive mesh},'' \emph{Philosophical Transactions of the Royal Society A:
  Mathematical, Physical and Engineering Sciences}, vol. 371, no. 1985, p.
  20120251, 2 2013. [Online]. Available:
  \url{https://royalsocietypublishing.org/doi/10.1098/rsta.2012.0251}
\BIBentrySTDinterwordspacing

\bibitem{Goss2019EconomicArrays}
Z.~L. Goss, S.~C. Kramer, A.~Avdis, C.~J. Cotter, and M.~D. Piggott,
  ``{Economic optimisation of large scale tidal stream turbine arrays},'' in
  \emph{13th European Wave and Tidal Energy Conference}, Naples, 2019, pp.
  1598--1598.

\bibitem{Goss2018AnRace}
\BIBentryALTinterwordspacing
Z.~L. Goss, M.~D. Piggott, and S.~C. Kramer, ``{An English Channel Model for
  the Optimisation of Tidal Turbines in the Alderney Race},'' in \emph{Oxford
  Tidal Energy Workshop}, Oxford, 2018, pp. 10--11. [Online]. Available:
  \url{http://www2.eng.ox.ac.uk/tidal/templates/OTE2018_proceedings_v3b.pdf}
\BIBentrySTDinterwordspacing

\bibitem{Goss2019VariationsCosts}
\BIBentryALTinterwordspacing
Z.~L. Goss, S.~C. Kramer, A.~Avdis, C.~J. Cotter, and M.~D. Piggott,
  ``{Variations in the optimal design of a tidal stream turbine array with
  costs},'' in \emph{Oxford Tidal Energy Workshop}, Oxford, 2019, pp. 33--34.
  [Online]. Available:
  \url{http://www2.eng.ox.ac.uk/tidal/ote2019-1/proceedings-ote2019}
\BIBentrySTDinterwordspacing

\bibitem{Culley2016TheArrays}
D.~M. Culley, S.~Funke, M.~Piggott, and A.~Peter, ``{The modelling and design
  optimisation of tidal stream turbine arrays},'' Ph.D. dissertation, Imperial
  College London, 2016.

\bibitem{Smart2018TidalBenefit}
\BIBentryALTinterwordspacing
G.~Smart and M.~Noonan, ``{Tidal stream and wave energy cost reduction and
  industrial benefit},'' \emph{Catapult ORE}, 2018. [Online]. Available:
  \url{https://www.marineenergywales.co.uk/wp-content/uploads/2018/05/ORE-Catapult-Tidal-Stream-and-Wave-Energy-Cost-Reduction-and-Ind-Benefit-FINAL-v03.02.pdf}
\BIBentrySTDinterwordspacing

\bibitem{Iyer2013VariabilityKingdom}
A.~S. Iyer, S.~J. Couch, G.~P. Harrison, and A.~R. Wallace, ``{Variability and
  phasing of tidal current energy around the United Kingdom},'' \emph{Renewable
  Energy}, vol.~51, pp. 343--357, 3 2013.

\bibitem{Thiebot2020InfluenceFrance}
\BIBentryALTinterwordspacing
J.~Thi{\'{e}}bot, S.~Guillou, and E.~Droniou, ``{Influence of the 18.6-year
  lunar nodal cycle on the tidal resource of the Alderney Race, France},''
  \emph{Applied Ocean Research}, vol.~97, p. 102107, 4 2020. [Online].
  Available: \url{doi:10.1016/j.apor.2020.102107}
\BIBentrySTDinterwordspacing

\bibitem{Haigh2011GlobalLevels}
\BIBentryALTinterwordspacing
I.~D. Haigh, M.~Eliot, and C.~Pattiaratchi, ``{Global influences of the 18.61
  year nodal cycle and 8.85 year cycle of lunar perigee on high tidal
  levels},'' \emph{Journal of Geophysical Research: Oceans}, vol. 116, no.~6, 6
  2011. [Online]. Available: \url{doi:10.1029/2010JC006645}
\BIBentrySTDinterwordspacing

\bibitem{SongThePerformance}
S.~Song, W.~Shi, Y.~K. Demirel, and M.~Atlar, ``{The effect of biofouling on
  the tidal turbine performance},'' Tech. Rep.

\bibitem{Faulstich2011WindDeployment}
\BIBentryALTinterwordspacing
S.~Faulstich, B.~Hahn, and P.~J. Tavner, ``{Wind turbine downtime and its
  importance for offshore deployment},'' \emph{Wind Energy}, vol.~14, no.~3,
  pp. 327--337, 4 2011. [Online]. Available: \url{doi:wiley.com/10.1002/we.421}
\BIBentrySTDinterwordspacing

\bibitem{Ziegler2018LifetimeUK}
\BIBentryALTinterwordspacing
L.~Ziegler, E.~Gonzalez, T.~Rubert, U.~Smolka, and J.~J. Melero, ``{Lifetime
  extension of onshore wind turbines: A review covering Germany, Spain,
  Denmark, and the UK},'' \emph{Renewable and Sustainable Energy Reviews},
  vol.~82, pp. 1261--1271, 2 2018. [Online]. Available:
  \url{doi:10.1016/j.rser.2017.09.100}
\BIBentrySTDinterwordspacing

\bibitem{Lewis2009RealRates}
\BIBentryALTinterwordspacing
N.~Lewis, T.~Eschenbach, and J.~Hartman, ``{Real Options And The Use Of
  Discrete And Continuous Interest Rates},'' \emph{American Society for
  Engineering Education}, no. 2009 Annual Conference {\&} Exposition, pp.
  1--14, 2009. [Online]. Available: \url{https://peer.asee.org/4987}
\BIBentrySTDinterwordspacing

\bibitem{Vazquez2015DeviceEnergy}
\BIBentryALTinterwordspacing
A.~Vazquez and G.~Iglesias, ``{Device interactions in reducing the cost of
  tidal stream energy},'' \emph{Energy Conversion and Management}, vol.~97, pp.
  428--438, 6 2015. [Online]. Available:
  \url{https://www.sciencedirect.com/science/article/pii/S0196890415002629#b0200}
\BIBentrySTDinterwordspacing

\bibitem{Vazquez2015LCOEEnergy}
\BIBentryALTinterwordspacing
------, ``{LCOE (levelised cost of energy) mapping: A new geospatial tool for
  tidal stream energy},'' \emph{Energy}, vol.~91, pp. 192--201, 11 2015.
  [Online]. Available:
  \url{https://www.sciencedirect.com/science/article/pii/S0360544215010828#bib21}
\BIBentrySTDinterwordspacing

\bibitem{Ouyang2014LevelizedChina}
\BIBentryALTinterwordspacing
X.~Ouyang and B.~Lin, ``{Levelized cost of electricity (LCOE) of renewable
  energies and required subsidies in China},'' \emph{Energy Policy}, vol.~70,
  pp. 64--73, 7 2014. [Online]. Available:
  \url{https://www.sciencedirect.com/science/article/pii/S0301421514001773}
\BIBentrySTDinterwordspacing

\bibitem{Vazquez2016CapitalApproach}
\BIBentryALTinterwordspacing
A.~Vazquez and G.~Iglesias, ``{Capital costs in tidal stream energy projects
  – A spatial approach},'' \emph{Energy}, vol. 107, pp. 215--226, 7 2016.
  [Online]. Available: \url{doi:10.1016/j.energy.2016.03.123}
\BIBentrySTDinterwordspacing

\bibitem{2020LessonsReport}
\BIBentryALTinterwordspacing
``{Lessons Learnt from MeyGen Phase 1A: Final Summary Report},'' 2020.
  [Online]. Available: \url{https://bit.ly/2BffaOD}
\BIBentrySTDinterwordspacing

\bibitem{2018MarineDevelopments}
\BIBentryALTinterwordspacing
``{Marine Scotland Review of Approaches and Cost of Decommissioning Small Scale
  Offshore Renewable Energy Developments},'' Tech. Rep., 2018. [Online].
  Available: \url{www.arup.com}
\BIBentrySTDinterwordspacing

\bibitem{Ahmadian2012Far-fieldTurbines}
R.~Ahmadian, R.~Falconer, and B.~Bockelmann-Evans, ``{Far-field modelling of
  the hydro-environmental impact of tidal stream turbines},'' \emph{Renewable
  Energy}, vol.~38, no.~1, pp. 107--116, 2 2012.

\bibitem{Ahmadian2012AssessmentOutput}
R.~Ahmadian and R.~A. Falconer, ``{Assessment of array shape of tidal stream
  turbines on hydro-environmental impacts and power output},'' \emph{Renewable
  Energy}, vol.~44, pp. 318--327, 8 2012.

\bibitem{MccluskieBirdsReview}
A.~E. Mccluskie, L.~R. H. W.~. Wilkinson, and N.~I. Rspb, \emph{{Birds and wave
  {\&} tidal stream energy: an ecological review}}.

\bibitem{duFeu2019TheApproach}
\BIBentryALTinterwordspacing
R.~J. du~Feu, S.~W. Funke, S.~C. Kramer, J.~Hill, and M.~D. Piggott, ``{The
  trade-off between tidal-turbine array yield and environmental impact: A
  habitat suitability modelling approach},'' \emph{Renewable Energy}, vol. 143,
  pp. 390--403, 12 2019. [Online]. Available:
  \url{doi:10.1016/j.renene.2019.04.141}
\BIBentrySTDinterwordspacing

\bibitem{Shields2011MarineEnvironment}
M.~A. Shields, D.~K. Woolf, E.~P. Grist, S.~A. Kerr, A.~C. Jackson, R.~E.
  Harris, M.~C. Bell, R.~Beharie, A.~Want, E.~Osalusi, S.~W. Gibb, and J.~Side,
  ``{Marine renewable energy: The ecological implications of altering the
  hydrodynamics of the marine environment},'' pp. 2--9, 1 2011.

\bibitem{Eckman1989EffectsSay}
J.~E. Eckman, C.~H. Peterson, and J.~A. Cahalan, ``{Effects of flow speed,
  turbulence, and orientation on growth of juvenile bay scallops Argopecten
  irradians concentricus (Say)},'' \emph{Journal of Experimental Marine Biology
  and Ecology}, vol. 132, no.~2, pp. 123--140, 11 1989.

\bibitem{Phillips2013MaxEnt}
\BIBentryALTinterwordspacing
R.~E.~S. Steven J.~Phillips, Miroslav~Dudík, ``{Maxent software for modeling
  species niches and distributions (Version 3.4.1)},'' 2018. [Online].
  Available: \url{http://biodiversityinformatics.amnh.org/open_source/maxent/}
\BIBentrySTDinterwordspacing

\bibitem{Nelson2017AConstraints}
\BIBentryALTinterwordspacing
K.~Nelson, S.~C. James, J.~D. Roberts, and C.~Jones, ``{A framework for
  determining improved placement of current energy converters subject to
  environmental constraints},'' \emph{International Journal of Sustainable
  Energy}, pp. 1--15, 6 2017. [Online]. Available:
  \url{https://www.tandfonline.com/doi/full/10.1080/14786451.2017.1334654}
\BIBentrySTDinterwordspacing

\bibitem{Ederer2015EvaluatingApproach}
N.~Ederer, ``{Evaluating capital and operating cost efficiency of offshore wind
  farms: A DEA approach},'' pp. 1034--1046, 2015.

\bibitem{Sullivan2013OptimisationAlgorithm}
\BIBentryALTinterwordspacing
P.~Sullivan and P.~McCombie, ``{Optimisation of tidal power arrays using a
  genetic algorithm},'' \emph{Proceedings of the Institution of Civil Engineers
  - Energy}, vol. 166, no.~1, pp. 19--28, 2 2013. [Online]. Available:
  \url{http://www.icevirtuallibrary.com/doi/10.1680/ener.12.00011}
\BIBentrySTDinterwordspacing

\bibitem{Schwedes2017MeshOptimisation}
T.~Schwedes, D.~A. Ham, S.~W. Funke, and M.~D. Piggott, \emph{{Mesh Dependence
  in PDE-Constrained Optimisation}}.\hskip 1em plus 0.5em minus 0.4em\relax
  Springer International Publishing, 2017.

\bibitem{Lewis2015ResourceArrays}
M.~Lewis, S.~P. Neill, P.~E. Robins, and M.~R. Hashemi, ``{Resource assessment
  for future generations of tidal-stream energy arrays},'' \emph{Energy},
  vol.~83, pp. 403--415, 4 2015.

\bibitem{Coles2019MechanismsEnergy}
D.~S. Coles and T.~Walsh, ``{Mechanisms for reducing the cost of tidal stream
  energy},'' in \emph{13th European Wave and Tidal Energy Conference}, Naples,
  2019, pp. 1836--1.

\bibitem{Arup2016Department}
\BIBentryALTinterwordspacing
O.~Arup, ``{Department of Energy and Climate Change Review of Renewable
  Electricity Generation Cost and Technical Assumptions Study Report Job number
  Job number Review of Renewable Electricity Generation Cost and Technical
  Assumptions Study Report Report Ref | Final |},'' Tech. Rep., 2016. [Online].
  Available: \url{www.arup.com}
\BIBentrySTDinterwordspacing

\bibitem{Ltd2017LessonsCONTENTS}
\BIBentryALTinterwordspacing
{Meygen Ltd}, ``{Lessons Learnt from MeyGen Phase 1a Part 1/3: Design Phase
  Lessons Learnt from MeyGen Phase 1a Part 1/3: Design Phase LESSONS LEARNT
  FROM MEYGEN PHASE 1A. PART 1/3: DESIGN PHASE CONTENTS},'' Tech. Rep., 2017.
  [Online]. Available:
  \url{https://tethys.pnnl.gov/sites/default/files/publications/MeyGen-2017-Part1.pdf}
\BIBentrySTDinterwordspacing

\bibitem{DynamicallyITPEnergised}
\BIBentryALTinterwordspacing
``{Dynamically Positioned Barge Set to Revolutionise Tidal Energy Installation
  - ITPEnergised}.'' [Online]. Available:
  \url{http://www.itpenergised.com/dynamically-positioned-barge-set-to-revolutionise-tidal-energy-installation/}
\BIBentrySTDinterwordspacing

\bibitem{2013OceanM}
\BIBentryALTinterwordspacing
``{Ocean Energy: Cost of Energy and Cost Reduction Opportunities M},'' 2013.
  [Online]. Available:
  \url{https://energiatalgud.ee/img_auth.php/1/10/SI_OCEAN._Ocean_Energy_-_Cost_of_Energy_and_Cost_Reduction._2013.pdf}
\BIBentrySTDinterwordspacing

\bibitem{Harcourt2019UtilisingValue}
F.~Harcourt, A.~Angeloudis, and M.~D. Piggott, ``{Utilising the flexible
  generation potential of tidal range power plants to optimise economic
  value},'' \emph{Applied Energy}, vol. 237, pp. 873--884, 3 2019.

\bibitem{Higgins2014TheKingdom}
P.~Higgins and A.~Foley, ``{The evolution of offshore wind power in the united
  kingdom},'' pp. 599--612, 9 2014.

\bibitem{Green2007ElectricalPower}
\BIBentryALTinterwordspacing
J.~Green, A.~Bowen, L.~J. Fingersh, and Y.~Wan, ``{Electrical Collection and
  Transmission Systems for Offshore Wind Power},'' Tech. Rep., 2007. [Online].
  Available: \url{http://www.osti.gov/bridge}
\BIBentrySTDinterwordspacing

\bibitem{Heptonstall2012TheFuture}
P.~Heptonstall, R.~Gross, P.~Greenacre, and T.~Cockerill, ``{The cost of
  offshore wind: Understanding the past and projecting the future},''
  \emph{Energy Policy}, vol.~41, pp. 815--821, 2 2012.

\bibitem{2015InternationalTechnology}
\BIBentryALTinterwordspacing
``{International LCOE for Ocean Energy Technology},'' 2015. [Online].
  Available:
  \url{https://www.ocean-energy-systems.org/news/international-lcoe-for-ocean-energy-technology/}
\BIBentrySTDinterwordspacing

\bibitem{MarkHamilton-CTO2012TechnologyUpdate}
{Mark Hamilton-CTO}, ``{Technology Update},'' Scotrenewables Tidal Power Ltd,
  Tech. Rep., 2012.

\bibitem{2010CostBackground}
``{Cost of and financial support for wave, tidal stream and tidal range
  generation in the UK Executive summary Background},'' Tech. Rep., 2010.

\bibitem{Khatib2010ReviewEdition}
H.~Khatib, ``{Review of OECD study into "Projected costs of generating
  electricity-2010 Edition"},'' \emph{Energy Policy}, vol.~38, no.~10, pp.
  5403--5408, 10 2010.

\bibitem{2006CostMethodology}
\BIBentryALTinterwordspacing
``{Cost estimation methodology},'' Tech. Rep., 2006. [Online]. Available:
  \url{https://www.carbontrust.com/media/54785/mec_cost_estimation_methodology_report.pdf}
\BIBentrySTDinterwordspacing

\bibitem{Allan2011LevelisedCertificates}
G.~Allan, M.~Gilmartin, P.~McGregor, and K.~Swales, ``{Levelised costs of Wave
  and Tidal energy in the UK: Cost competitiveness and the importance of
  "banded" renewables obligation certificates},'' \emph{Energy Policy},
  vol.~39, no.~1, pp. 23--39, 1 2011.

\bibitem{IEANEA2015Projected2015}
{IEA NEA}, \emph{{Projected Costs of Generating Electricity 2015}}, 2015.

\bibitem{2016ELECTRICITYCOSTS}
\BIBentryALTinterwordspacing
``{ELECTRICITY GENERATION COSTS},'' Tech. Rep., 2016. [Online]. Available:
  \url{www.nationalarchives.gov.uk/doc/open-government-licence/}
\BIBentrySTDinterwordspacing

\bibitem{Dalton2015EconomicPerspectives}
\BIBentryALTinterwordspacing
G.~Dalton, G.~Allan, N.~Beaumont, A.~Georgakaki, N.~Hacking, T.~Hooper,
  S.~Kerr, A.~Marie~O'hagan, K.~Reilly, P.~Ricci, W.~Sheng, and T.~Stallard,
  ``{Economic and socio-economic assessment methods for ocean renewable energy:
  Public and private perspectives},'' 2015. [Online]. Available:
  \url{http://dx.doi.org/10.1016/j.rser.2015.01.068}
\BIBentrySTDinterwordspacing

\bibitem{Klaus2020FinancialStudy}
\BIBentryALTinterwordspacing
S.~Klaus, ``{Financial and Economic Assessment of Tidal Stream Energy—A Case
  Study},'' \emph{International Journal of Financial Studies}, vol.~8, no.~3,
  p.~48, 8 2020. [Online]. Available:
  \url{https://www.mdpi.com/2227-7072/8/3/48}
\BIBentrySTDinterwordspacing

\bibitem{MeyGenEnergy}
\BIBentryALTinterwordspacing
``{MeyGen Phase 1A completes construction phase and officially enters 25 year
  operations phase | SIMEC Atlantis Energy}.'' [Online]. Available:
  \url{https://simecatlantis.com/2018/04/12/meygen-phase-1a-completes-construction-phase-and-officially-enters-25-year-operations-phase/}
\BIBentrySTDinterwordspacing

\bibitem{Johnstone2013ATechnology}
C.~M. Johnstone, D.~Pratt, J.~A. Clarke, and A.~D. Grant, ``{A techno-economic
  analysis of tidal energy technology},'' \emph{Renewable Energy}, vol.~49, pp.
  101--106, 1 2013.

\bibitem{Mills2019Impact2017}
A.~D. Mills, D.~Millstein, R.~Wiser, J.~Seel, J.~P. Carvallo, S.~Jeong, and
  W.~Gorman, ``{Impact of Wind, Solar, and Other Factors on Wholesale Power
  Prices An Historical Analysis-2008 through 2017},'' Tech. Rep., 2019.

\bibitem{Neill2014OptimalFunction}
S.~P. Neill, M.~R. Hashemi, and M.~J. Lewis, ``{Optimal phasing of the European
  tidal stream resource using the greedy algorithm with penalty function},''
  \emph{Energy}, vol.~73, pp. 997--1006, 8 2014.

\bibitem{2019UKIndustry}
\BIBentryALTinterwordspacing
``{UK Marine Energy 2019 - A new industry},'' Marine Energy Council,
  RenewableUK and Scottish Renewables, Tech. Rep., 2019. [Online]. Available:
  \url{https://www.marineenergywales.co.uk/wp-content/uploads/2019/03/uk_marine_energy_2019.pdf}
\BIBentrySTDinterwordspacing

\bibitem{Conroy2011WindIreland}
N.~Conroy, J.~P. Deane, and B.~P. O~Gallachoir, ``{Wind turbine availability:
  Should it be time or energy based? - A case study in Ireland},''
  \emph{Renewable Energy}, vol.~36, no.~11, pp. 2967--2971, 11 2011.

\bibitem{2017DefinitionsIndustry}
``{Definitions of Availability Terms for the Wind Industry},'' Tech. Rep.,
  2017.

\bibitem{2017LessonsPhase}
\BIBentryALTinterwordspacing
``{Lessons Learnt from MeyGen Phase 1a Part 1/3: Design Phase},'' 2017.
  [Online]. Available:
  \url{https://tethys.pnnl.gov/sites/default/files/publications/MeyGen-2017-Part1.pdf}
\BIBentrySTDinterwordspacing

\end{thebibliography}

\end{document}